\documentclass[12pt]{article}

\usepackage{amsmath}
\usepackage{graphicx}
\usepackage{enumerate}
\usepackage{natbib}
\usepackage{url} 
\usepackage{amssymb}
\usepackage{amsbsy}
\usepackage{amsfonts,mathrsfs,moreverb,lipsum}
\usepackage{graphicx,colonequals}
\usepackage{fixltx2e}
\usepackage{color}
\usepackage{caption}
\usepackage{subcaption}
\usepackage{yfonts}
\usepackage{pbox}
\usepackage{booktabs}
\usepackage{array}
\usepackage[pdfencoding=auto]{hyperref}

\newcommand{\blind}{1}

\newcommand{\mL}{\mathcal{L}}

\newcommand{\pig}{\pi_g}

\newcommand{\vecD}{\mathbf{D}}

\newcommand{\vecx}{\mathbf{x}}
\newcommand{\vecy}{\mathbf{y}}
\newcommand{\vecu}{\mathbf{u}}
\newcommand{\vecX}{\mathbf{X}}

\newcommand{\vecV}{\mathbf{V}}

\newcommand{\vecz}{\mathbf{z}}
\newcommand{\vecp}{\mathbf{p}}

\newcommand{\matlam}{\mathbf\Lambda}
\newcommand{\matdel}{\mathbf\Delta}

\newcommand{\matSig}{\mathbf\Sigma}

\newcommand{\matphi}{\mathbf\Phi}

\newcommand{\vecGam}{\mbox{\boldmath$\Gamma$}}

\newcommand{\varthet}{\mbox{\boldmath$\vartheta$}}

\newcommand{\tr}{\,\mbox{tr}}

\newcommand{\vecA}{\mathbf{A}}

\newcommand{\matt}{\mathbf{T}}
\newcommand{\matd}{\mathbf{D}}

\newcommand{\tenM}{\textgoth{M}}

\newcommand{\tenX}{\textgoth{X}}
\newcommand{\tenV}{\textgoth{V}}
\newcommand{\ident}{\mathbf{I}}

\newcommand{\veckappa}{\mbox{\boldmath$\kappa$}}

\newcommand{\vecDelta}{\mbox{\boldmath$\Delta$}}
\newcommand{\vecPhi}{\mbox{\boldmath$\Phi$}}

\newcommand{\fX}{\mathscr{X}}

\newcommand{\vecc}{\text{vec}}

\newcommand{\mhalf}{-\frac{1}{2}}
\newcommand{\T}{^{\top}}
\newcommand{\inv}{^{-1}}

\newcommand{\vecn}{\mathbf{n}}
\newcommand{\vecM}{\mathbf{M}}

\newcommand{\cholDeltaI}[1]{\vecDelta_{#1}^{\mhalf}}

\newcommand{\norm}[1]{\left\lVert#1\right\rVert}
\newcommand{\vecXbrev}{\breve{\vecX}}

\newcommand{\pd}[1]{\frac{\partial}{\partial #1}}

\newcommand{\evec}[2]{\mathbf{e}^{#1}_{#2}} 

\begin{document}

\def\spacingset#1{\renewcommand{\baselinestretch}%
{#1}\small\normalsize} \spacingset{1}


\if1\blind
{
  \title{\bf Clustering Higher Order Data:\\ An Application to Pediatric Multi-variable Longitudinal Data}
  \author{Peter A. Tait and  Paul D. McNicholas\\ 
    Dept.\ of Mathematics \& Statistics, McMaster University, Ontario, Canada.\\
    Joyce Obeid \\
    Dept.\ of Pediatrics , McMaster University, Ontario, Canada.\thanks{
	The authors gratefully acknowledge a grant-in-aid from the Heart and Stroke Foundation of Canada (G-14-0005722) as well as funding from the Canada Research Chairs program and an E.W.R.\ Steacie Memorial Fellowship from the Natural Sciences and Engineering Research Council of Canada.}\hspace{.2cm}\\}
  \maketitle
} \fi

\if0\blind
{
  \bigskip
  \bigskip
  \bigskip
  \begin{center}
    {\LARGE\bf Title}
\end{center}
  \medskip
} \fi

\bigskip
\begin{abstract}
	Physical activity levels are an important predictor of cardiovascular health and increasingly being measured by sensors, like accelerometers. Accelerometers produce rich multivariate data that can inform important clinical decisions related to individual patients and public health. The CHAMPION study, a study of youth with chronic inflammatory conditions, aims to determine the links between heart health, inflammation, physical activity, and fitness. The accelerometer data from CHAMPION is represented as $4$-dimensional arrays, and a finite mixture of multidimensional arrays model is developed for clustering. The use of model-based clustering for multidimensional arrays has thus far been limited to two-dimensional arrays, i.e., matrices or order-two tensors, and the work in this paper can also be seen as an approach for clustering $D$-dimensional arrays for $D>2$ or, in other words, for clustering order-$D$ tensors.
\end{abstract}

\noindent%
\vfill

\newpage
\spacingset{1.5} 
\section{Introduction}
\label{sec:intro}

Regular physical activity and a high cardiovascular fitness, as measured by VO2max, are strong independent predictors of a persons health status in both children and adults \citep{who2010}. The world health organization recommends a minimum of sixty minutes of moderate or vigorous physical activity per day and muscle strengthening activities at least three times a week. Large epidemiological studies indicate that there is a strong relationship between being physically active and a decreased risk of cardiovascular disease \citep[CVD;][]{manson2002, yu2003}, a leading cause of mortality in the developed world. It is estimated that only 33\% of children meet these physical activity guidelines \citep{colley2019} and as a result, many are potentially at risk of developing CVD. Across age groups, physical activity is strongly related to aerobic fitness. In children, high aerobic fitness has been linked to lower body mass index, blood pressure, cholesterol and is considered and important means of lowering a child's risk of CVD \citep{carnethon2003}.

Children growing up with a medical condition can be at higher risk than their healthy counterparts for developing CVD because of their impaired ability to meet the physical activity guidelines \citep{kavey2006}. This impairment can be due to their medical condition, other lifestyle factors or an interaction of the two. The CHAMPION study investigates heart health, as well as the lifestyle factors that might affect it, in the most common chronic diseases of childhood. The study includes a group of healthy children as a control. Accelerometers are used to capture the physical activity levels of the children. We represent the accelerometer data as a sample of five-way data, where each patient has a four-dimensional tensor of data. As is common in biostatistics, the accelerometer data is longitudinal and spans multiple time scales (e.g. secs, minutes, hours), each of which could have different patterns of variation that offer important scientific insights. Using this depiction of the data, we can capture all these patterns of variation and the mean trends across the time scales simultaneously with one statistical model.

Cluster analysis, or clustering, has long played an important role as an exploratory technique for multivariate data from a wide variety of sources \citep[see][]{everitt11}. We apply clustering to the CHAMPION data to separate the study population into homogeneous subgroups. These subgroups have differing physical activity profiles, as measured by the accelerometers. In addition to highlighting differences in the variation and mean activity among the subgroups, the cluster labels can be used as low dimensional summaries of the five-way data in statistical models relating them to measures of heart health. A sizeable proportion of the statistical literature uses finite mixture models \citep[see, e.g., reviews by][]{fraley02a,bouveyron14,mcnicholas16b} to cluster multivariate (i.e., two-way) data. With the increasing size and complexity of modern data, there is a need for clustering approaches that are effective for higher order data. Recently, there have been some notable examples of clustering three-way data using finite mixtures of matrix-variate distributions \citep[e.g.,][]{viroli2011, anderlucci2015, gallaugher18}. While this work on clustering three-way data is timely, there is no reason to stop at three-way data.

\section{Background}

\subsection{Multidimensional Arrays}

An approach for clustering multi-way data is introduced based on a finite mixture of multidimensional arrays. While some might refer to such structures as `tensors', and so write about clustering tensor-variate data, we prefer the nomenclature multidimensional array (MDA) to avoid confusion with the term `tensor' as used in engineering and physics, e.g., tensor fields. Consider data arranged in a MDA. Herein, and on account of the data at hand, we restrict ourselves to data that can be regarded as the realization of continuous random variables. The number of dimensions a tensor has is referred to as its order. An order-$D$ tensor is equivalent, in our sense, to a $D$-dimensional array --- the $D=2$ structure is a matrix, the $D=3$ structure can be regarded as a rectangular cuboid. A rectangular cuboid is defined as a three-dimensional box, where all the angles are right angles, all faces are rectangles, and opposite faces are equal \citep{harris1998}. A $D=4$ structure can be viewed as stacked rectangular cuboids. MDAs can be partitioned into slices or matrices which are two-dimensional sections of a MDA. This is done by fixing all but two dimensions of the MDA \citep{kolda2009}.  In general, $(D+1)$-way data can be represented using a sample of $D$-dimensional MDAs. 

\subsection{Multilinear Normal Distribution}
The multilinear normal distribution can be used to model $D$-dimensional MDAs that are populated with continuous data. It has been described in detail by \cite{ohlson2013}. If $\mathcal{X}$ is a random order-$D$ MDA, with dimensional lengths $\vecn = [n_1,\ldots,n_D]$, with realization $\tenX$, then it follows that the probability density function of a multilinear normal distribution (MLND), $\mathcal{N}_{n^{*}}(\tenM,\bigotimes_{d=1}^D\vecDelta_d)$, can be written
\begin{align}
\label{eq:tensnorm}
\nonumber 
f_{\text{\tiny MLND}}\bigg(\tenX|&\tenM, \bigotimes_{d=1}^D\vecDelta_d \bigg) = f_{\text{\tiny MLND}}(\tenX|\tenM, \vecDelta_1,\ldots,\vecDelta_D) = \\
& (2\pi)^{\frac{n^{*}}{2}}\prod_{d=1}^D|\vecDelta_d|^{-\frac{n^{*}}{2n_d}} \exp\left\{-\frac{1}{2}\vecc(\tenX-\tenM)\T\bigotimes_{d=1}^D\vecDelta_d^{-1}\vecc(\tenX-\tenM)\right\}\text{,}
\end{align}

where $\vecc(\cdot)$ is the vectorization operator, $\tenM$ is the mean MDA, $n^{*} = \prod_{d=1}^Dn_d$,  $\bigotimes_{d=1}^D\vecDelta_d = \vecDelta_1 \otimes \dots \otimes \vecDelta_D$, $\otimes$ denotes the Kronecker product, and $\vecDelta_d$ is a scale matrix of dimension~$d$. 

\subsection{Finite Mixture Models}\label{finite mixture model}

The term model-based clustering is often used to describe clustering based on finite mixture models. A random variable $\vecX$ follows a finite mixture model if its density is a convex linear combination of probability densities, i.e., if, for all $\vecx\subset\vecX$, its density can be written
\begin{equation}\label{eq:finite mixture model}
f(\vecx~|~{\varkappa})=\sum_{g=1}^{G}\pig f_{g}(\vecx~|~{\boldsymbol \theta}_g),
\end{equation}
where the $\pi_g > 0$ is the $g$th mixing proportion with $\sum_{g=1}^{G}\pig=1$, $f_{g}(\vecx~|~{\boldsymbol \theta}_g)$ is the $g$th component density with parameters ${\boldsymbol \theta}_g$, and ${\varkappa}=\{\pi_g,{\boldsymbol \theta}_g\}_{g=1}^G$ denotes all parameters of the mixture model. Detailed accounts of finite mixture models and their application for clustering can be found in \cite{mclachlan2004finite} and \cite{mcnicholas2016}. 

Consider a mixture of order-$d$ dimensional arrays. Specifically, the density of $\mathcal{X}$ following a mixture of MLNDs is
\begin{equation}
f(\tenX~|~\varthet) = \sum^G_{g=1}\pig f_{\text{\tiny MLND}}(\tenX|\tenM_g, \vecDelta_{g,1},\ldots,\vecDelta_{g,D}) \text{,}
\end{equation}
where $\varthet = \{\pi_g, \tenM_g, \vecDelta_{g,1},\ldots,\vecDelta_{g,D}\}_{g=1}^G$ represents the parameters of this mixture of MLNDs, each of which has the same interpretation as before.
One can interpret $\pi_g$ as the \textit{a~priori} probability that the $i$th array $\tenX_i$ belongs to the $g$th component of the mixture.

\section{Methods}
\label{sec:meth}

\subsection{Matricizations}

For ease of exposition, we focus on order-$3$ MDAs, which are cuboids. Suppose there are $1,\dots,n_1$ rows, $1,\dots,n_2$ columns, and $1,\dots,n_3$ frontal slices.  Let $\tenX = (x_{ijk}): n_1 \times n_2 \times n_3$ be a three-dimensional array. The vectorization of the array $\tenX$ is defined as follows
\begin{equation}
\vecc(\tenX) = \sum_{i = 1}^{n_1} \sum_{j = 1}^{n_2}\sum_{k = 1}^{n_3} x_{ijk}\evec{1}{i}\otimes\evec{2}{j}\otimes\evec{3}{k},
\end{equation} 
where $x_{ijk}$ is ${(i,j,k)}$th element of $\tenX$ and  $\evec{1}{i}$, $\evec{2}{j}$ and $\evec{3}{k}$ are unit basis vectors of size $n_1$, $n_2$ and $n_3$, respectively. The mode one matricization of $\tenX$, $\vecX_{(1)}$ can be expressed as 
\begin{equation*}
\vecX_{(1)} = \sum_{i = 1}^{n_1} \sum_{j = 1}^{n_2}\sum_{k = 1}^{n_3} x_{ijk}(\evec{2}{j}\otimes\evec{3}{k})\evec{1\top}{i}.
\end{equation*}
Noting that $\vecc(\vecx\vecy\T) = \vecy \otimes \vecx$, we have $\vecc(\tenX) = \vecc(\vecX_{(1)})$. This result generalizes to order-$D$ arrays, which have $D$ unit basis vectors.

For parameter estimation, the exponent in \eqref{eq:tensnorm} can be rewritten as
\begin{align}
\nonumber \vecc(\tenX-\tenM)\T\bigotimes_{d=1}^D\vecDelta_d^{-1}\vecc(\tenX-\tenM) &= 
\vecc(\vecX_{(1)}-\vecM_{(1)})\T \vecc\left(\bigotimes_{d=2}^D\vecDelta_d^{-1}(\vecX_{(1)}-\vecM_{(1)})\vecDelta_1^{-1}\right)\\
&=\tr\left[\vecDelta_1^{-1}(\vecX_{(1)}-\vecM_{(1)})\T\bigotimes_{d=2}^D\vecDelta_{d}^{-1}(\vecX_{(1)}-\vecM_{(1)})\right], \label{eq:trace1}
\end{align}
where $\vecV_{(1)}$ is the mode $1$ matricization of the MDA $\tenV$. This formulation leads to the following equivalence, 
$$
\fX\sim \mathcal{N}_{\vecp}\left(\tenM,\bigotimes_{d=1}^D\vecDelta_d\right) \iff \fX_{(j)}\sim \mathcal{N}_{\frac{n^*}{n_1}\times n_1}\left(\vecM_{(1)},\bigotimes_{d=2}^D\vecDelta_d, \vecDelta_1\right),
$$
where $\mathcal{N}_{\frac{n^*}{n_1}\times n_1}(\cdot)$ denotes the matrix variate normal distribution with ${n^*}/{n_1}$ rows and $n_1$ columns, mean matrix $\vecM_{(1)}$ and scale matrices $\vecDelta_1$ and $\bigotimes_{d=2}^D\vecDelta_d$.

The entire trace in \eqref{eq:trace1} can be further decomposed by noting that $n^{*}_{3:D} = \prod_{d=3}^Dn_d$, $\ident_{n^{*}_{3:d}} = \sum_{j=1}^{n^{*}_{3:D}}\evec{}{j}\evec{\top}{j}$, $\cholDeltaI{d}$ is the Cholesky decomposition of $\vecDelta_d\inv$, and $\vecXbrev_{(1)} = \vecX_{(1)} - \vecM_{(1)}$:
\begin{align}
\nonumber
\tr\left[\vecDelta_1^{-1}\vecXbrev_{(1)}\T\bigotimes_{d=2}^D\vecDelta_{d}^{-1}\vecXbrev_{(1)}\right] &= \tr\left[\vecDelta_1^{-1}\vecXbrev_{(1)}\T\vecDelta_{2}^{-1}\otimes \bigotimes_{d=3}^D\vecDelta_{d}^{-1}\vecXbrev_{(1)}\right]\\
\label{eq:trace2}
&=\sum_{j=1}^{n^{*}_{3:D}} \tr\left[\vecDelta_1\inv \vecX_{(1), j}\T \vecDelta_{2}\inv \vecX_{(1), j}\right],
\end{align}
where $\vecX_{(1), j} = \left(\ident_{n_2} \otimes \mathbf{e}_{j}\T \bigotimes_{d=3}^D\cholDeltaI{d}\right)\vecXbrev_{(1)}$. See appendix \ref{app:A} for details.

We use the tensor commutation operators, outlined in \cite{ohlson2013}, to permute the rows of the mode one matricizations and entries of the Kronecker products. It exchanges the second and $l^{th}$ elements in the sequence of unit basis vectors or scale matrices, where $3 \leq l \leq D$. We denote these modifications by the superscript $l,2$. Equation \eqref{eq:trace2} can be re-expressed as 
\begin{equation}\label{eq:trace3}
\sum_{j=1}^{n^{*}_{2:D/l}} \tr\left[\vecDelta_1\inv (\vecX_{(1),j}^{l,2})\T \vecDelta_l\inv \vecX_{(1),j}^{l,2}\right],
\end{equation}
where $n^{*}_{2:D/l} = \prod_{\substack{d=2\\d\ne l}}^D n_d$ and $\vecX_{(1),j}^{l,2} = \bigg(\ident_{n_l} \otimes \evec{\top}{j}\bigotimes_{\substack{d=2\\d\ne l}}^D \cholDeltaI{d} \bigg)\vecXbrev_{(1)}^{l,2}$.


\subsection{Likelihoods}

Given a sample of $N$ iid random $D$-dimensional arrays $\tenX_1, \dots, \tenX_N$, the complete-data likelihood is given by
\begin{equation*}
\mL_{C}(\varthet) = \prod^G_{g=1}\prod^N_{i=1}[\pig f_{\text{\tiny MLND}}(\tenX_i~|~\tenM_g, \vecDelta_{g,1},\ldots,\vecDelta_{g,D})]^{z_{ig}},
\end{equation*}
where  $z_{ig}=1$ if $\tenX_i$ belongs to component $g$ and $z_{ig}=0$ otherwise. Note that $\vecz_i=(z_{i1},\ldots,z_{iG})'$ is considered a realization of a multinomial random variable with one draw on $G$ categories and probabilities given by $\pi_1,\ldots,\pi_G$.
Taking the natural logarithm of the complete-data likelihood,
\begin{equation}\label{eq:ll1}\begin{split}
\log\mL_{C}(\varthet) &= \sum^G_{g=1}\sum^N_{i=1}z_{ig}[\log\pig + \log f_{\text{\tiny MLND}}(\tenX_i~|~\tenM_g, \vecDelta_{g,1},\ldots,\vecDelta_{g,D})]\\
 =&  C + \sum^G_{g=1}n_g\log\pig -\frac{n^*}{2}\sum^G_{g=1}n_g\sum_{d=1}^D\frac{1}{n_d}\log(|\vecDelta_{g,d}|)\\ 
& -\frac{1}{2}\sum^G_{g=1}\sum_{i=1}^N z_{ig}\tr\left[\vecDelta_{g,1}^{-1}\vecXbrev_{(1), g,i}\T\bigotimes_{d=2}^D\vecDelta_{d}^{-1}\vecXbrev_{(1),g,i}\right],
\end{split}\end{equation}
where $C$ is a constant that does not depend on the parameters and $n_g = \sum^N_{i=1}z_{ig}$. 

Depending on the permutation of the MDA we use, the final term in \eqref{eq:ll1} can be replaced with  \eqref{eq:trace2} or \eqref{eq:trace3} as
\begin{equation}\label{eq:ll2}
-\frac{1}{2}\sum^G_{g=1}\sum^N_{i=1}z_{ig}\sum_{j=1}^{n^{*}_{3:D}} \tr\left[\vecDelta_{g,1}\inv \vecX_{(1),g,i,j}\T \vecDelta_{g,2}\inv \vecX_{(1),g,i,j}\right]
\end{equation}
or
\begin{equation}\label{eq:ll3}
-\frac{1}{2}\sum^G_{g=1}\sum^N_{i=1}z_{ig} \sum_{j=1}^{n^{*}_{2:D/l}} \tr\left[\vecDelta_{g,1}\inv (\vecX_{(1),g,i,j}^{l,2})\T \vecDelta_{g,l}\inv \vecX_{(1),g,i,j}^{l,2}\right],
\end{equation}
resulting in three complete-data log-likelihood equations. These equations allow us to isolate the individual model parameters and enable their estimation.  

\subsection{Parameter Estimation}
\subsubsection{E- and M-steps}\label{sec:vvv}
The parameters of the models describe herein are all estimated by the method of maximum likelihood. The maximum likelihood estimates are found using the expectation-maximization (EM) algorithm, a two-step iterative algorithm used to calculated the parameter estimates in the presence of missing data \citep{dempster1977}. In the E-step, the expected value of the complete-data log-likelihood, $\mathcal{Q}$, is computed conditional on the current parameter estimates. In the M-step, $\mathcal{Q}$ is maximized resulting in new, or updated, parameter estimates. The E- and M-steps are iterated until some stopping rule is satisfied (Section~\ref{sec:stopping}).

In our case, the E-step amounts to replacing the $z_{ig}$ values in \eqref{eq:ll1} by their conditional expectations
\begin{equation}\label{eq:estep}
\hat{z}_{ig} = \frac{\hat{\pi}_g f_{\text{\tiny MLND}}(\tenX_i~|~\tenM_g, \vecDelta_{g,1},\ldots,\vecDelta_{g,D})}{\sum^G_{h=1}\hat{\pi}_h f_{\text{\tiny MLND}}(\tenX_i~|~\tenM_h, \vecDelta_{h,1},\ldots,\vecDelta_{h,D})}.
\end{equation} 
The M-step update for $\varthet$ are available in closed form and essentially follow from taking respective first derivatives of $\mathcal{Q}$ and setting the resulting expressions to zero. 
The update for $\pi_g$ is given by $\hat{\pi}_g={n_g}/{N}$. The respective M-step updates for $\vecDelta_{g,1}$ and $\vecDelta_{g,2}$ involve taking the first derivative of the $\mathcal{Q}$ function that uses \eqref{eq:ll2} as its final term. The updates are
\begin{equation}\label{eq:md1}
\hat{\vecDelta}_{g,1} = \frac{n_1}{n^{*} n_g} \sum^N_{i=1} \hat{z}_{i,g} \sum_{j=1}^{n^{*}_{3:D}} \vecX_{(1),g,i,j}\T \vecDelta_{g,2}\inv \vecX_{(1),g,i,j}
\end{equation}
and 
\begin{equation}\label{eq:md2}
\hat{\vecDelta}_{g,2} = \frac{n_2}{n^{*} n_g} \sum^N_{i=1} \hat{z}_{i,g} \sum_{j=1}^{n^{*}_{3:D}} \vecX_{(1),g,i,j} \vecDelta_{g,1}\inv \vecX_{(1),g,i,j}\T,
\end{equation}
respectively. The update for $\vecDelta_{g,l}$ uses the $\mathcal{Q}$ function that adopts \eqref{eq:ll3} for its final term. The update is 
\begin{equation}\label{eq:md3}
\hat{\vecDelta}_{g,l} = \frac{n_l}{n^{*} n_g} \sum^N_{i=1} \hat{z}_{i,g} \sum_{j=1}^{n^{*}_{2:D/l}} \vecX_{(1),g,i,j}^{l,2} \vecDelta_{g,1}\inv (\vecX_{(1),g,i,j}^{l,2})\T.
\end{equation}
The M-step update for $\vecM_{(1)}$, uses the $\mathcal{Q}$ function equivalent to \eqref{eq:ll1} and is given by
\begin{equation}\label{eq:mmean}
\hat{\vecM}_{(1),g} = \frac{1}{n_g} \sum^N_{i=1}\hat{z}_{i,g}\vecX_{(1),i}.
\end{equation}

\subsubsection{Stopping Rule}\label{sec:stopping}

To stop our EM algorithms, we use a criterion based on the Aitken acceleration \citep{aitken26}. At iteration $t$ of the EM algorithm, the Aitken acceleration is 
\begin{equation}
a^{(t)}=\frac{l^{(t+1)}-l^{(t)}}{l^{(t)}-l^{(t-1)}},
\end{equation}
where $l^{(t)}$ is the (observed) log-likelihood at iteration $t$. \cite{bohning94} use $a^{(t)}$ to calculate an asymptotic estimate of the log-likelihood at iteration $t+1$:
\begin{equation}
l^{(t+1)}_\infty=l^{(t)}+\frac{1}{1-a^{(t)}}(l^{(t+1)}-l^{(t)}).
\end{equation}
We stop the EM algorithm when $l^{(t+1)}_\infty-l^{(t)}<\epsilon$ \citep{lindsay1995, mcnicholas10a}.

\subsection{Model Selection}

The number of groups $G$ in a clustering problem is often unknown \textit{a priori}. In such cases, the parameters of a mixture model are typically estimated for different values of $G$ and some criterion is then used to select~$G$. 
The most common choice for this criteria is the Bayesian information criteria \citep[BIC;][]{schwarz78}, which can be written
\begin{align}
\text{BIC} = 2 l(\hat{\varthet})- \rho \log N \text{,}
\end{align}
where $l(\hat{\varthet})$ is the maximized log-likelihood, $\rho$ is the number of free parameters in the model and $N$ is the number of observations. 
For our finite mixture of multilinear normal distributions, 
\begin{align}\label{eq:fp}
\rho = (G-1) + Gn^{*} + \dfrac{G}{2} \sum_{d=1}^{D} n_d(n_d+1).
\end{align}

\subsection{Identifiability}\label{sec:ident}

The scale parameters in the Kronecker product are unique up to a strictly positive multiplicative constant. Indeed, if we let $d_k > 0$ then  
\begin{equation}\label{eq:identif}
\bigotimes_{d=1}^D\vecDelta_d = \frac{1}{\prod_{k=2}^{D}d_k}\vecDelta_1 \otimes \bigotimes_{k=2}^D d_k\vecDelta_k,  
\end{equation}
and the likelihood is unchanged. However, we notice that, $\bigotimes_{d=1}^D\vecDelta_d = \bigotimes_{d=1}^D\tilde{\vecDelta}_d$, where $\tilde{\vecDelta}_d$ are the terms on the right hand side of \eqref{eq:identif}, so the estimate of the Kronecker product would be unique. This constraint is imposed once the EM algorithm has converged. 
There are two well known options for specifying the values of the $d_k$. The first is letting $d_k = 1/\vecDelta_k(1,1)$, where $\vecDelta_k(1,1)$ is the first entry in $\vecDelta_k$ and the second is setting $d_k = {n_k}/{\tr\vecDelta_k}$, which equates to setting $\tr\vecDelta_k = n_k$ \citep{anderlucci2015}. We use the former option for $d_k$ in our models.

\subsection{Parsimony} 
\subsubsection{Families of Mixture Models}
The number of free parameters, given by \eqref{eq:fp}, can be substantial as $D$ and the values in $\vecn$ increase. We denote this model VVV. It is described in Section~\ref{sec:vvv} and does not have any parsimony constraints imposed on its parameters. Most of the parameters are accounted for by the $\vecDelta_d$ matrices, which are a natural place to impose parsimony constraints. In practice, often one or more dimensions of the MDA are composed of ordered values, usually some representation of time. We use the modified Cholesky decomposition \citep[MCD;][]{pourahmadi1999} to constrain the number of parameters in the $\vecDelta_d$ modeling temporal dimensions of the MDAs. We extend two members of the eight member Cholesky-decomposed Gaussian mixture model family \citep[CDGMM;][]{mcnicholas2010model} to our mixture model. We chose the VVI and EVI models because they have either variable or equal autoregressive coefficients between time points for the groups. They both include an isotropic constraint on the  variability at each time point.    

The Cholesky decomposition \citep{benoit1924} decomposes a square positive definite matrix into a unique lower triangular matrix and its transpose, see \cite{Golub2013} for a detailed discussion. The MCD reorganizes the standard Cholesky decomposition into a diagonal matrix $\matd$ and triangular matrix $\matt$ with ones on the main diagonal. The MCD decomposition of the precision matrix $\vecDelta_{d}\inv$ can be expressed as:
\begin{equation*}
\vecDelta_{d}\inv = \matt_d\T \matd_d\inv \matt_d.
\end{equation*}  
The isotropic constraint for $\matd_d$ is expressed as $\matd_d = \delta_d\ident_{n_d}$, where $\matd_d$ is $n_d \times n_d$.

For non-temporal dimensions of the MDA, we implement the VVI and EEE models, two members of the 14 Gaussian parsimonious clustering models \citep[GPCM;][]{celeux1995}. These models constrain the eigen-decomposition of the associated scale matrix. The decomposition has the following form:
\begin{equation*}
\vecDelta_d = \lambda_d\vecGam_d\vecD_d\vecGam_d\T,
\end{equation*}
where  $\lambda_d=|\vecDelta_d|^\frac{1}{n_d}$, $\vecD_g$ is a diagonal matrix containing the normalized eigenvalues of $\vecDelta_d$ in decreasing order and $\vecGam_d$ is the corresponding matrix of eigenvectors of $\vecDelta_d$. This decomposition can be interpreted geometrically, where $\lambda_d$ is the volume, $\vecD_d$ is the shape, and $\vecGam_d$ is the orientation of the component. 

The EEE model assumes that each group has the same scale structure and, of all the GPCM constraints with $n_d^2$ parameters, is the one with the fewest. The VVI model assumes a diagonal parameterization of the scale matrix where the cluster orientations are axis aligned but have varying volume and shape parameters. It has $n_d$ free parameters. It was chosen because it has the most general parameterization with $n_d$ parameters and could give a hint towards the importance of modeling the complete variation in each dimension of the MDA data. 

\subsubsection{CDGMM Parameter Estimation}

To apply the VVI model to $\vecDelta_{1, g}$, we start with \eqref{eq:ll2} and re-express the terms related to $\vecDelta_{1, g}$ as:  
\begin{equation}\label{eq:vvi_d1}
\log\mL_{C}(\varthet) = C + \frac{n^*}{2}\sum^G_{g=1}n_g\log(\delta_{g,1}\inv) -\frac{1}{2}\sum^G_{g=1}n_g\delta_{g,1}\inv\tr\big[\matt_{g,1}\matlam_{g,1}\matt_{g,1}\T\big],
\end{equation} 
where $\matlam_{g,1} = \frac{1}{n_g}\sum_{i=1}^N \hat{z}_{ig}\sum_{j=1}^{n^{*}_{3:D}}\vecX_{(1),g,i,j}\T\matdel_{g,2}\inv\vecX_{(1),g,i,j}$.
After taking the partial derivatives of the $\mathcal{Q}$ function associated with \eqref{eq:vvi_d1}, we end up with the following score functions:
\begin{align}
\label{eq:vvi_s1}
\text{S}_1(\delta_{g,1}, \matt_{g,1}) &= \frac{n_g}{2}(n^{*}\delta_{g,1} - \tr[\matt_{g,1}\matlam_{g,1}\matt_{g,1}\T]),\\
\label{eq:vvi_s2}
\text{S}_2(\delta_{g,1}, \matt_{g,1}) &= -\frac{n_g}{\delta_{g,1}} \matt_{g,1}\matlam_{g,1}.
\end{align}
For details, see Appendix \ref{app:B}. The first score function can be solved for $\delta_{g,1}$, resulting in the following expression: 
\begin{equation}
\delta_{g,1} = \frac{1}{n^{*}}\tr[\matt_{g,1}\matlam_{g,1}\matt_{g,1}\T].
\end{equation}

Given we are only interested in estimating the lower triangular components of $\matt_{g,1}$ that are below the main diagonal, we can designate these elements as $\vecPhi_{g,1}$ which is a lower triangular matrix with ${n_1(n_1-1)}/{2}$ elements to be estimated. If you substitute $\vecPhi_{g,1}$ for $\matt_{g,1}$ into \eqref{eq:vvi_s2} and set it equal to zero, it can be shown that for $r = 2,\dots n_1$ we have:
\[
\left[ {\begin{array}{c}
	\phi^{(g),1}_{r,1} \\
	\phi^{(g),1}_{r,2} \\
	\vdots \\
	\phi^{(g),1}_{r,r-1}	
	\end{array} } \right] = -
\left[ {\begin{array}{cccc}
	\lambda^{(g),1}_{11} & \lambda^{(g),1}_{21} & \dots & \lambda^{(g),1}_{r-1,1} \\
	\lambda^{(g),1}_{12} & \lambda^{(g),1}_{22} & \dots & \lambda^{(g),1}_{r-1,2} \\
	\vdots & \vdots & \ddots & \vdots \\
	\lambda^{(g),1}_{1,r-1} & \lambda^{(g),1}_{2, r-1} & \dots & \lambda^{(g),1}_{r-1, r-1}
	\end{array} } \right]\inv
\left[ {\begin{array}{c}
	\lambda^{(g),1}_{r,1} \\
	\lambda^{(g),1}_{r,2} \\
	\vdots \\
	\lambda^{(g),1}_{r,r-1}	
	\end{array} } \right].
\]
This is equivalent to solving the following system of equations for $\matphi^{(g),1}_{(r-1) \times 1}$: 
\begin{equation*}
\matlam^{(g),1 \top}_{(r-1) \times (r-1)}\matphi^{(g),1}_{(r-1) \times 1} = - \matlam^{(g),1}_{(r-1) \times 1}.
\end{equation*}
The VVI model for $\vecDelta_{1, g}$ has $G{n_1(n_1-1)}/{2} + G$ free parameters. 

To find the VVI models for $\matdel_{g,2}$ and $\matdel_{g,l}$, we follow the same steps outlined above but we start with \eqref{eq:ll2} and \eqref{eq:ll3}, respectively. The new $\matlam$ matrices are defined as follows:
\begin{align*}
\matlam_{g,2} &= \frac{1}{n_g}\sum_{i=1}^N \hat{z}_{ig}\sum_{j=1}^{n^{*}_{3:D}}\vecX_{(1),g,i,j}\matdel_{g,1}\inv\vecX_{(1),g,i,j}\T\\
\matlam_{g,l} &= \frac{1}{n_g}\sum_{i=1}^N \hat{z}_{ig}\sum_{j=1}^{n^{*}_{2:D/l}} \vecX_{(1),g,i,j}^{l,2}\matdel_{g,1}\inv(\vecX_{(1),g,i,j}^{l,2})\T,
\end{align*}
and the results follow as outlined above.

To apply the EVI model to $\vecDelta_{1, g}$, we start with \eqref{eq:ll2} and re-express the terms related to $\vecDelta_{1, g}$ as:  
\begin{equation*}\label{eq:evi_d1}
\log\mL_{C}(\varthet) = C + \frac{n^*}{2}\sum^G_{g=1}n_g\log(\delta_{g,1}\inv) -\frac{1}{2}\sum^G_{g=1}n_g\delta_{g,1}\inv\tr\big[\matt_1\matlam_{g,1}\matt_1\T\big],
\end{equation*} 
where $\matlam_{g,1}$ is defined as above. 
The score functions are:
\begin{align}
\text{S}_1(\delta_{g,1}, \matt_1) &= \frac{n_g}{2}( n^*\delta_{g,1} - \tr[\matt_1\matlam_{g,1}\matt_1\T])\\
\text{S}_2(\delta_{g,1}, \matt_1) &= -\matt_1\sum^G_{g=1}\frac{n_g}{\delta_{g,1}}\matlam_{g,1}
\end{align}
The derivations are similar to the ones outlined in appendix \ref{app:B} for the VVI model. The associated expression for $\delta_{g,1}$ is: 
\begin{equation}
\delta_{g,1} = \frac{1}{n^*}\tr[\matt_1\matlam_{g,1}\matt_1\T]
\end{equation}
Similar to the VVI model, we are interested in the lower triangular elements of $\matt_1$ and denote them as the $\veckappa_1$ matrix. If we define the elements of $\veckappa_1$ as $\kappa_{i,j}^1 = \sum^G_{g=1}\frac{n_g}{\delta_{g,1}}\lambda_{1,i,j}^{(g)}$, we can solve the following system of equations for $\veckappa_1$ when $r = 2,\dots n_1$:
\begin{equation*}
\veckappa^{1, \top}_{(r-1) \times (r-1)}\matphi^1_{(r-1) \times 1} = - \veckappa^1_{(r-1) \times 1}.
\end{equation*}
The EVI model for $\vecDelta_{1, g}$ has ${n_1(n_1-1)}/{2} + G$ free parameters. 

Following similar lines, the EVI models can be found for $\matdel_{g,2}$ and $\matdel_{g,l}$ using the appropriate $\matlam_g$ matrices defined above.

\subsubsection{GPCM Parameter Estimation}

Starting with \eqref{eq:ll2}, an EEE model for $\matdel_{g, 2}$ can be formulated by re-organize the equation in terms of $\matdel_{g, 2}$ as follows: 
\begin{equation*}
\log\mL_{C}(\varthet) = C -\frac{1}{2}\left(\frac{n^*N}{n_2}\log(|\matdel_2|) + \tr\left[\sum^G_{g=1}n_g\matlam_{g,2}\matdel_2\inv\right] \right),
\end{equation*}
where $\matlam_{g,2}$ is defined above. Using Theorem~A.2 from \cite{celeux1995}, we find that 
\begin{equation}
\matdel_2 = \frac{n_2}{n^*N} \sum^G_{g=1}n_g\matlam_{g,2}
\end{equation}
The number of free parameters for the EEE model of $\matdel_{g, 2}$ is ${n_2(n_2+1)}/{2}$. The same derivation can be done for $\matdel_{g, 1}$ and $\matdel_{g, l}$ by substituting the appropriate $\matlam_g$ matrix into \eqref{eq:ll2} or \eqref{eq:ll3}.

The VVI model for $\matdel_{g, 2}$ represents the scale matrix as $\lambda_{g,2}\vecD_{g,2}$. Reorganizing \eqref{eq:ll2}, we have:
\begin{equation*}
\log\mL_{C}(\varthet) = C -\frac{1}{2}\bigg(n^*\sum^G_{g=1}n_g\log(\lambda_{g,2}) + \sum^G_{g=1}n_g(\lambda_{g,2}\inv)\tr\big[\matlam_{g,2}\vecD_{g,2}\inv\big] \bigg)
\end{equation*}
Using the Corollary A.1 from \cite{celeux1995}, $\vecD_{g,2}$ is given by:
\begin{align}
\vecD_{g,2} &= \frac{\text{diag}(\matlam_{g,2})}{|\text{diag}(\matlam_{g,2})|^{\frac{1}{n_2}}},\\
\lambda_g &= \frac{n_2}{n^*}|\text{diag}(\matlam_{g,2})|^{\frac{1}{n_2}}.
\end{align}
The number of free parameters for the VVI model of $\matdel_{g, 2}$ is $Gn_2$.

\subsection{Implementation}\label{sec:imp}

We have used version 1.3.0 of the Julia language \citep{julia17, mcnictait2019} to implement the finite mixture of MLNDs described herein. Typically data scientists choose the {\sf R} language \citep{Rcitation} to implement their algorithms. We have deviated from the norm here because, for our purposes, Julia offers a number of advantages over {\sf R}. Julia is a language specifically designed for numerical computing. It is a dynamic language which checks data types, modifies objects and functions at run-time and not compile time, making the user's programming experience similar to {\sf R}. Despite this, it has performance approaching statically typed languages such as C and FORTRAN. Julia is perhaps a more desirable choice than {\sf R} when fast run time speed is required, when implementing an algorithm from scratch, or when a parallel or distributed implementation is desired. 

We used Algorithm 4.1 from \cite{blanchard2019} to implement the log-sum-exp rule. Singular $\vecDelta_{g}$ values were numerically regularized by adding a small positive quantity to the diagonal elements of the matrices \citep{williams2006}. The regularization is summarized in the following equation:
\begin{equation}\label{eq:covreg}
\tilde{\vecDelta} = \hat{\vecDelta} + \epsilon\ident,
\end{equation}
where $\epsilon \in (0,0.1]$, $\hat{\vecDelta}$ is the estimated singular scale matrix, and $\tilde{\vecDelta}$ is the regularized estimate of $\vecDelta$. We used $\epsilon = 0.001$ in our implementation. The singularity of $\hat{\vecDelta}$ was assessed by checking if its inverse condition number is less than machine epsilon. The positive definiteness of the $\vecDelta_d$ matrices was checked with the Cholesky decomposition. The figures herein we made using Julia, using version 1.2.1 of the {\tt Gadfly} visualization package \citep{gadfly}.

\section{Simulation}

We conducted a simulation study to investigate the effect of different sample and MDA sizes on the following questions. Can we effectively estimate the model parameters, can we effectively capture the original group labels? How often do singular scale matrices occur and how do singular scale matrices affect the model results? Accurately estimating and interpreting the model parameters and labels are common goals of clustering in data science applications and as such, important to assess for new models. 

We used five-way data and the VVV model in our simulations. The simulations were conducted for sample sizes (e.g., $N$) of 60, 90, 120 and 180 subjects with three equal sized groups. The $n^*$ quantity was used to measure the different dimensions of the order four MDAs. Its values included 256, 625, 1296 and 2401. While these values of $n^*$ can equate to any product of dimension lengths, the simplest way to visualize the resulting MDAs is as an order 4 MDA with four equal dimension lengths of 4, 5, 6 or 7. For each combination of $N$ and $n^*$, 250 simulations were conducted. We took advantage of the Julia's native distributed computing capabilities to make these simulations computationally feasible.

Following definition 2.2 in \cite{ohlson2013}, the simulated data was generated using the following equation:  
\begin{equation}\label{eq:sim_mln}
\vecc(\tenX) = \vecc(\tenM) + \bigotimes_{d=1}^D\vecDelta_d^{\frac{1}{2}}\vecu \text{,}
\end{equation}
where $\vecu$ is a vector of iid $\mathcal{N}(0,1)$ random numbers. This is equivalent to the multivarite normal model for the vectorized version of the MDA data. This formulation has the disadvantage of having to recreate the tensor from the vectorized data and dealing with a potentially large matrix, $\bigotimes_{d=1}^D\vecDelta_d^{\frac{1}{2}}$. If we generate $\vecu$ and $\tenM$  as a mode $d$ MDAs, we can use tensor $d$-mode products to implement the final term on the right hand side of \eqref{eq:sim_mln} \citep{kolda2009}.  A tensor $d$-mode product multiplies an MDA by a matrix in mode $d$ and has the advantage of retaining the mode $d$ structure of the data and not creating one large matrix from the Kronecker product, $\bigotimes_{d=1}^D\vecDelta_d^{\frac{1}{2}}$ and then having to permute the data back into an MDA.


A signal-to-noise ratio of one was applied to the simulated data prior to analysis. The $\vecDelta_d$ parameters were generated by specifying a diagonal matrix of eigenvalues and a random orthogonal matrix and combining them as you would in an eigen-decomposition of the scale matrix. An $n_d \times n_d$ orthogonal matrix was created by generating $n_d^2$ iid $N(0,1)$ random values, placing them in a matrix and orthogonalizing it with the QR decomposition. We restrict the condition number of these $\vecDelta_d$ matrices to be at most 10.

The EM algorithm used to generate the results was initialized with identity matrices for all the scale matrix parameters. The values of $\hat{z}_{ig}$ are initialized with $k$-means starts, calculated on the $\vecc(\tenX_i)$ version of the data. The BIC was used to select the number of groups in our simulation. We checked $G \in \{2,3,4,5\}$ for each parameterization and in every instance, $G=3$ was chosen.

We use the relative error to determine how close the estimated model parameters were to the true parameters. It is defined as follows:
\begin{equation*}
\text{Relative Error} = \frac{\norm{\hat{\vecV} - \vecV}_F}{\norm{\vecV}_F},
\end{equation*}
where $\norm{*}_F$ is the Frobenius matrix norm, $\hat{\vecV}$ is the estimated parameter value, and $\vecV$ is the true parameter value used to generate the simulated data. The smaller this ratio is, the less error is present in the model's parameter estimates.
\begin{figure}
	\begin{center}
		\includegraphics[width=3.55in]{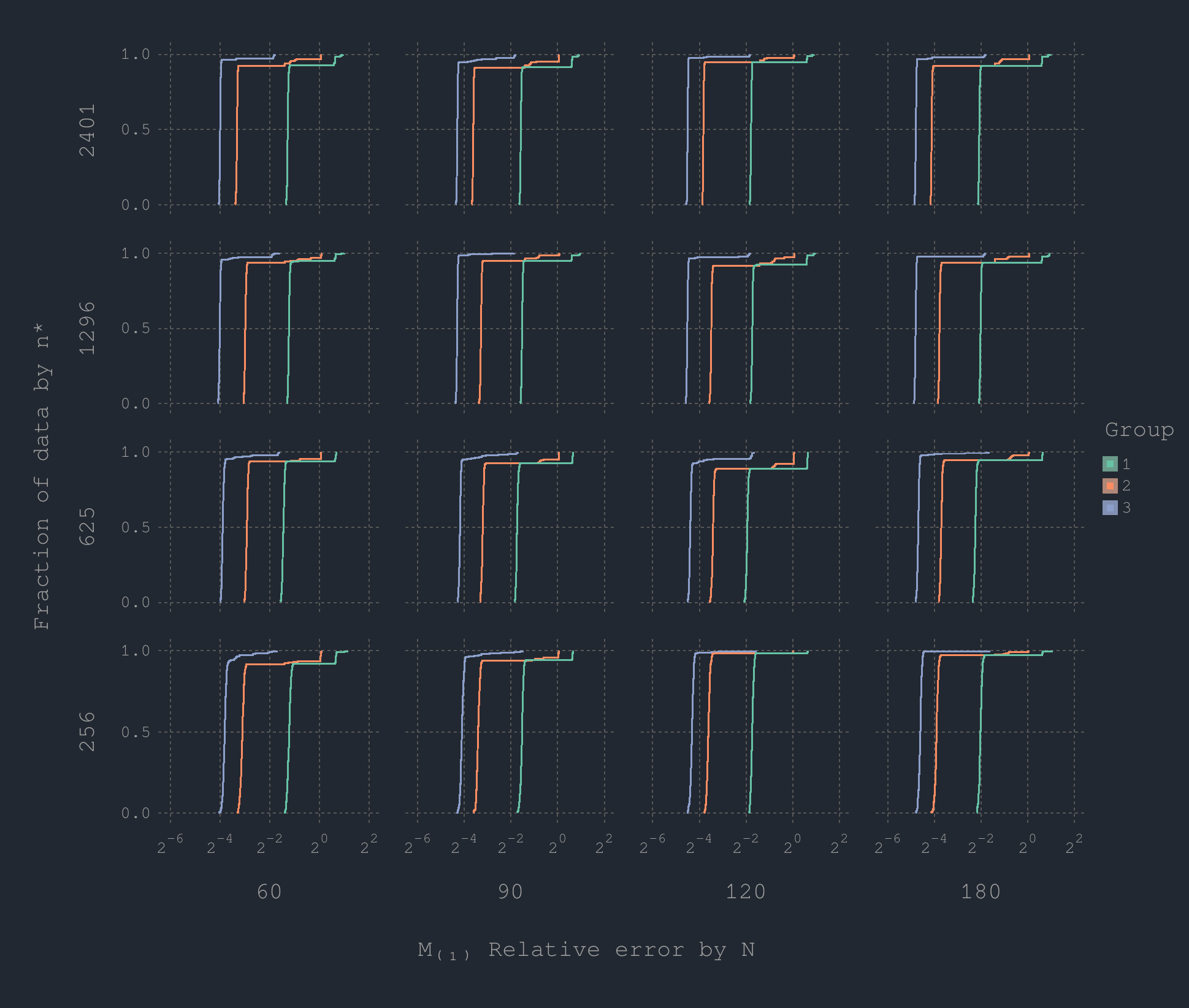}
		\caption{Relative errors for $\vecM_{(1),g}$ matrices.}
		\label{fig:sim_mean}
	\end{center}
\end{figure}

We use empirical cumulative distribution plots to summarize the distribution of the relative errors across the simulations for each value of $N$ and $n^*$. Figure \ref{fig:sim_mean} illustrates the relative error of the models $\vecM_{(1),g}$ parameter. The x-axis is plotted on the log2 scale because the distributions have a very long right tail. For the three groups, over $95\%$ of the values are well below one, indicating the mean matrices are being estimated accurately, with $\hat{\vecM}_{(1),1}$ consistently being estimated the least accurately. 

\begin{figure}[h]
\begin{center}
	\includegraphics[width=3.55in]{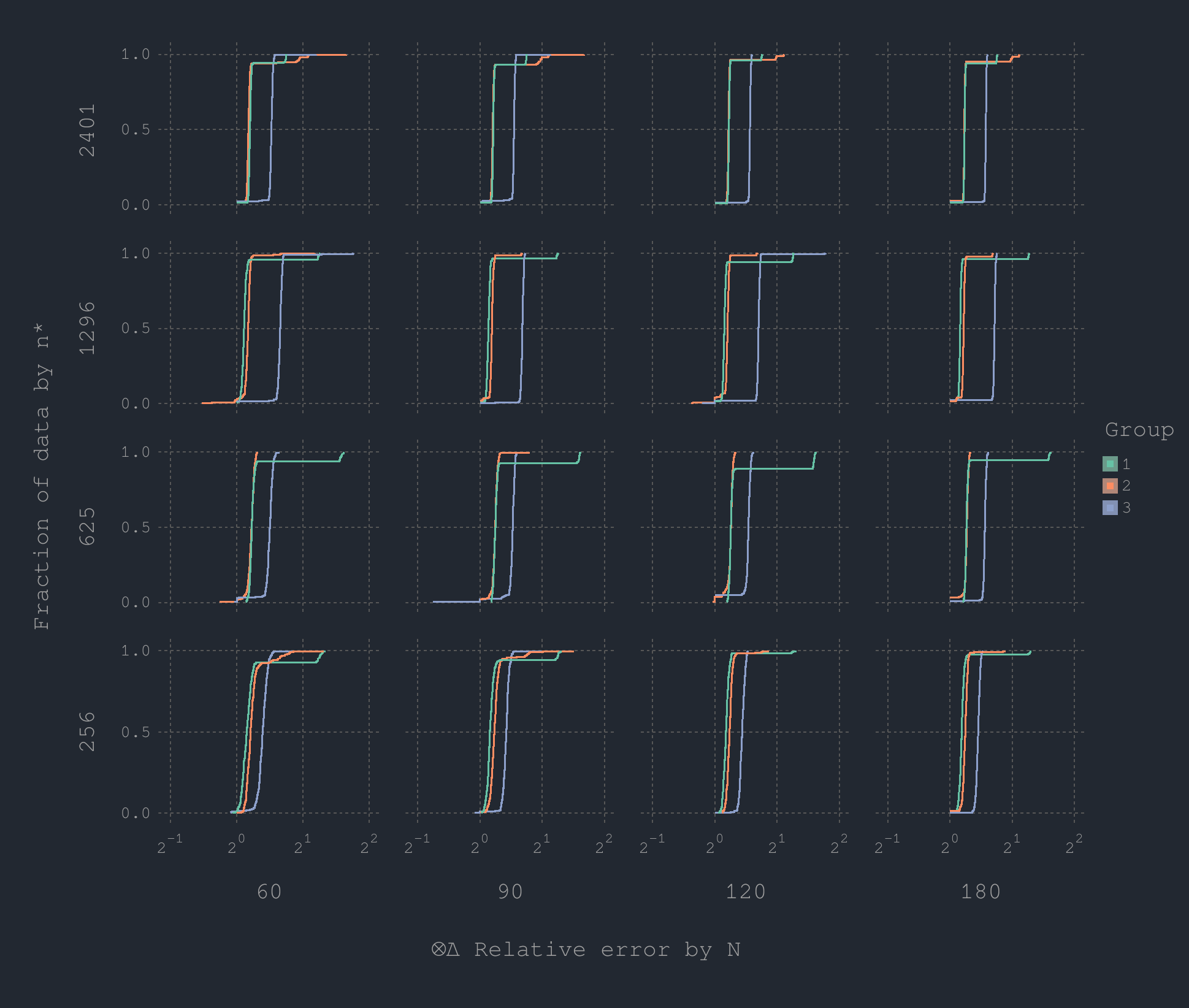}
	\caption{Relative errors for $\bigotimes_{d=1}^D\vecDelta_{d, g}$.}
	\label{fig:sim_kp}
\end{center}
\end{figure}

As noted in Section~\ref{sec:ident}, we would expect to be able to accurately estimate $\bigotimes_{d=1}^D\vecDelta_{g,d}$. Figure~\ref{fig:sim_kp} shows the distribution of the relative error for this quantity. The relative errors are larger than $\vecM_{(1),g}$ but still consistently between 1 and 1.5 for all combinations of $N$ and $n^*$. Of the three groups, $\bigotimes_{d=1}^D\hat{\vecDelta}_{3,d}$ is estimated the least accurately.

The group labels produced by the finite mixture model were compared to the simulated group labels via the adjusted Rand index \citep[ARI;][]{hubert85}. The Rand index is the ratio of the pair agreements to the total number of pairs \citep{rand71}. Chance agreement can enlarge the RI, making it difficult to interpret. The ARI corrects the Rand index for chance; it has an expected value of zero under random classification and retains the property of being equal to 1 when there is perfect class agreement. The average ARI for each combination of $N$ and $n^*$ was at least $0.95$, with an overall average of $0.969$ and standard deviation of $0.118$.
\begin{figure}[h]
	\begin{center}
		\includegraphics[width=3.55in]{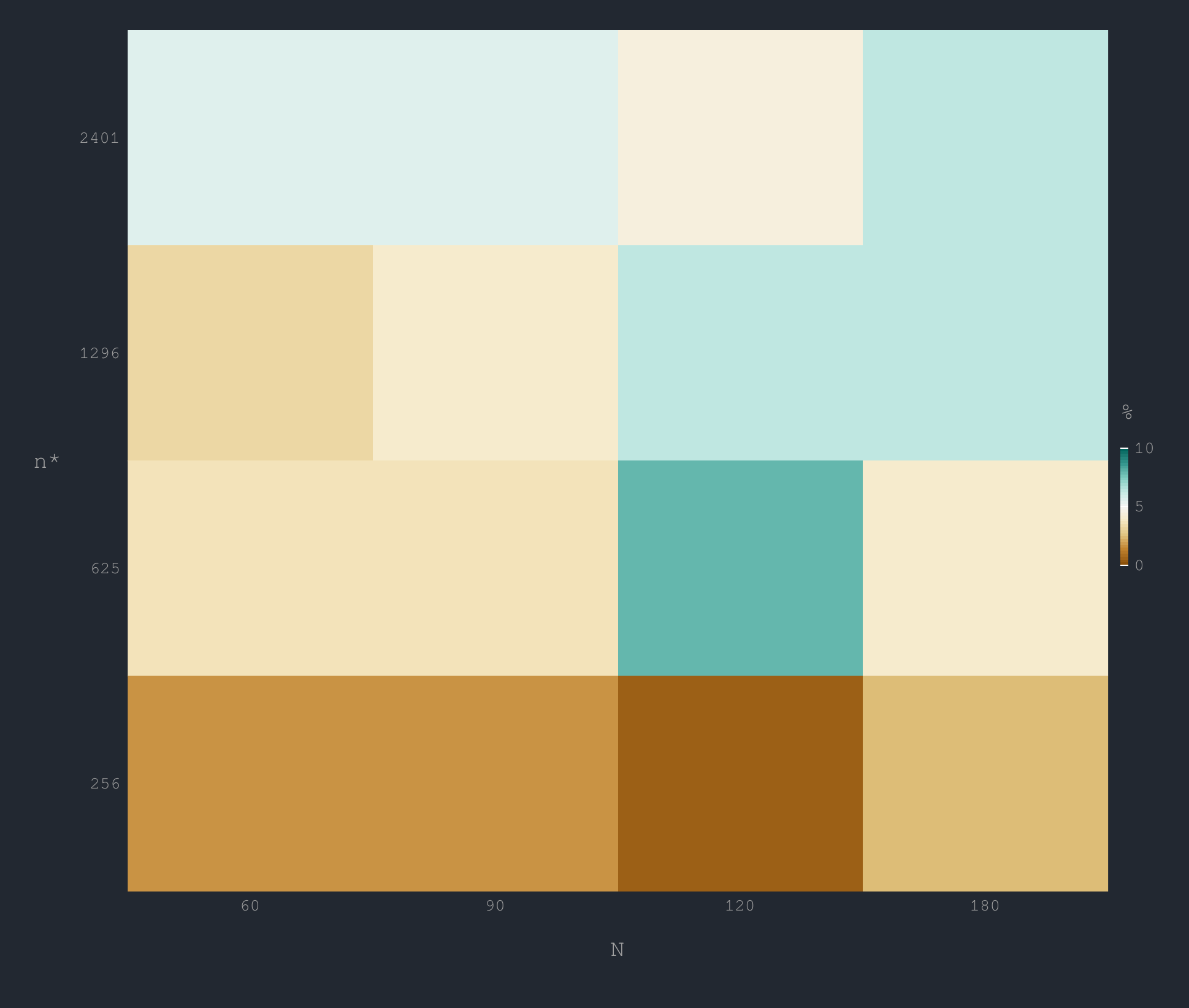}
	\caption{Percent of total simulations with a singular $\vecDelta_{d, g}$.}
	\label{fig:sim_sing}
\end{center}
\end{figure}

Given the high-dimensional data being analyzed and the large number of parameters being estimated by our model, we expected the ``curse of dimensionality'' to be a problem. In multivariate mixture models, if $N \ll p$, where $p$ is the number of variables being clustered, the scale matrices, $\matSig_{g, (p \times p)}$ can become become ill-conditioned and inverting them can lead to unstable results. This is some-what attenuated in our model because we are estimating individual $\vecDelta_{d}$, which are much lower dimensional than $\bigotimes_{d=1}^D\vecDelta_{d}$ and each $n_d$ is less than $N$. Nevertheless, we count the number of simulations that singular $\vecDelta_{g,d}$ occur and investigate how their occurrence affects the results.

Figure \ref{fig:sim_sing} is a heatmap of the percentage of the 250 simulations where a singular $\vecDelta_{g,d}$ occurred. When one occurred, it typically happened at the first iteration of the EM algorithm and would occur in a single group. No more than $10\%$ of the simulations had singular $\vecDelta_{d}$ matrices The smallest MDAs tended to have the least singular scale matrices and large values of $N$ and $n^*$ resulted in the most. Prior to conducting the simulations, we expected that small $N$ and large $n^*$ values would have had more singular scale matrices. 

\begin{figure}
	\begin{center}
		\includegraphics[width=3.55in]{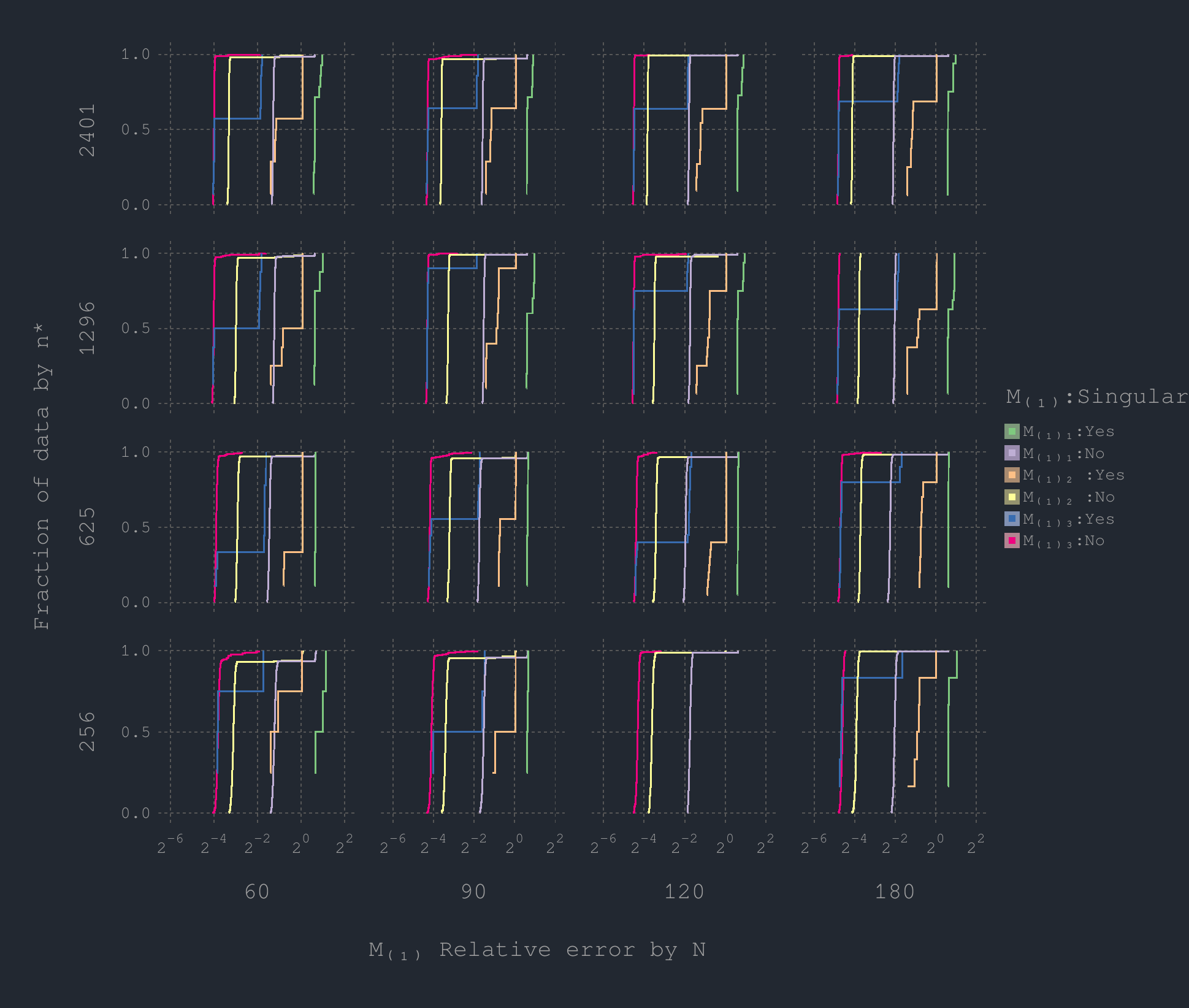}
	\caption{Relative errors for $\vecM_{(1),g}$ matrices by singular occurrence.}
	\label{fig:sim_mean_s}
\end{center}
\end{figure}

Figure \ref{fig:sim_mean_s} displays the relative error of $M_{(1),g}$ stratified by the occurrence of a singular $\vecDelta_{g,d}$. Uniformly, the estimates when singular $\vecDelta_{g,d}$ occur are less accurate. Interestingly, even the worst errors are still marginally larger than one, suggesting mean matricizations can still be reliably estimated.

\begin{figure}
	\begin{center}
		\includegraphics[width=3.55in]{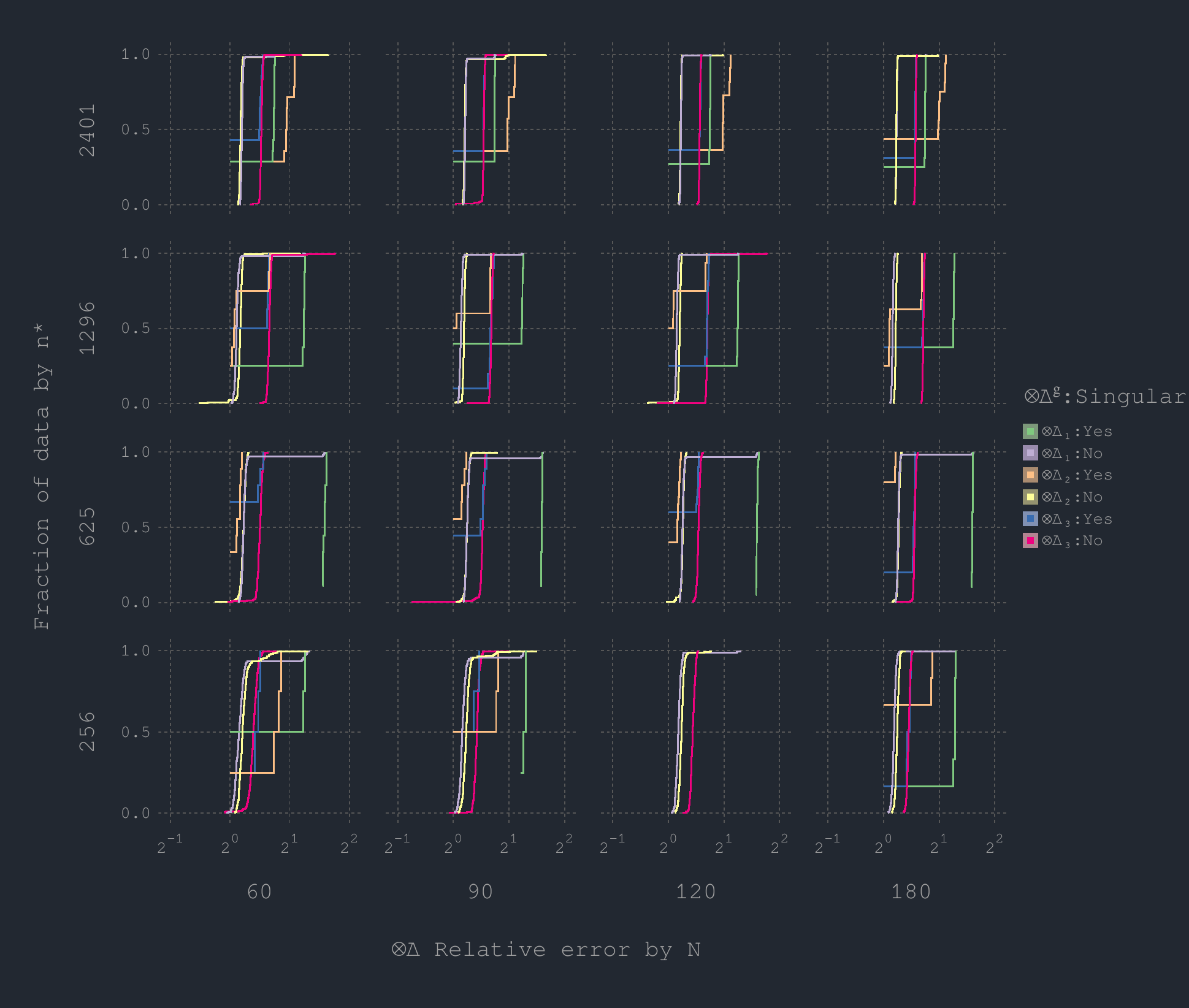}
	\caption{Relative errors for $\bigotimes_{d=1}^D\vecDelta_{d, g}$ by singular occurrence.}
	\label{fig:sim_kp_s}
\end{center}
\end{figure}

Figure \ref{fig:sim_kp_s}, illustrates the the relative errors of $\bigotimes_{d=1}^D\vecDelta_{d, g}$ stratified by the occurrence of a singular $\vecDelta_{g,d}$. The estimates without singular scale estimates are very close to one with $\vecDelta_{d, 3}$ consistently having the least accuracy. When singular $\vecDelta_{d, g}$'s occur, the overall errors are closer to 2 with group 1 having the worst results in low to moderate $n^*$ values and group 2 at the largest value of $n^*$. The mean ARI values are strongly affected by the singular $\vecDelta_{g,d}$'s. The results are summarized in table \ref{tab:sim_ari}. The summary statistics are calculated across all the values of $N$ and $n^*$. 
\begin{table}[h!]
	\centering
	\caption{ARI summaries for models with and without singular $\vecDelta_{g,d}$.}\label{tab:sim_ari}
	\begin{tabular}{lrrr}
		\hline
		\textbf{Metric} & \textbf{N} & \textbf{Mean} & \textbf{SD} \\ \hline
		Overall         & 4000 & 0.969 & 0.118 \\
		Singular: Yes   & 168 & 0.559 & 0.008 \\
		Singular: No    & 3822 & 0.987 & 0.080 \\ \hline
	\end{tabular}
\end{table} 

When singular $\vecDelta_{g,d}$ occur, the ARI and estimates of $\vecDelta_{g,d}$ are adversely affected, making their interpretation suspect.

\section{CHAMPION Study}\label{sec:champ}

The CHAMPION (Cardiovascular Health in children with a chronic inflAMmatory condition: role of Physical activity, fItness, and inflammatiON) study included youth between the ages of 7 and 17 years with a single diagnosis of a chronic inflammatory condition (CIC) including chronic cystic fibrosis (CF), juvenile idiopathic arthritis (JIA), inflammatory bowel disease (IBD), or type 1 diabetes mellitus (T1D) recruited from the McMaster Children's Hospital. These youths were required to have a confirmed diagnosis for at least 1 year, no medication changes (dosage or type) in the last month, and no hospitalizations for the last 3 months. Healthy control participants were recruited from the general community and were required to have no suspected or confirmed medical conditions and no medication usage. The goal of the CHAMPION study is to study the factors that might affect heart health in the most common chronic diseases of childhood.

CHAMPION is a cross-sectional, observational study where each participant was outfitted with an ActiGraph GT3X accelerometer (ActiGraph Corporation, Pensacola, USA), which has been used widely in the pediatric population to objectively quantify habitual physical activity and sedentary time \citep{evenson2008, colley2011, walker2015, stephens2016}. The ActiGraph is a microelectromechanical capacitive accelerometer that captures accelerations in the  vertical (axis1), anterioposterior (axis2), and mediolateral (axis3) planes at user-selected sampling frequencies. For the current study, raw accelerations were collected at 30 Hz, then filtered and converted into activity counts averaged over 3-sec sampling intervals, as previously described \citep{obeid2011, obeid2014, obeid2015, walker2015}.  The three axis values were combined to form the vector magnitude (VM), which is defined as $\sqrt{\text{axis1}^2 + \text{axis2}^2 +\text{axis3}^2}$. Participants were instructed to wear the device over the right hip for seven days during all waking hours, with the exception of water activities. 

A sample of 83 participants' accelerometer data was analyzed. They had a median of seven wear days. Because youth tend to exhibit short bursts of activity through out the day, understanding intraday activity patterns may have important implications for health. With this in mind, we aggregated the accelerometer data across each participants wear days into the following nested time periods; four 15 second, six 10 minute and 12 one hour periods (9h--20h). This results in 288 unique epochs per participant. Because the accelerometer data is all $\ge 0$ and heavily right skewed, we used the square-root transformation to transform the data before aggregation. The activity counts and VM were aggregated across days by taking their mean values. The participants steps were summed within each day and time period, the transformation was applied and the values were averaged across days. We included the standard deviation (SD) of VM and steps, calculated at the same time as the mean, to capture the variation in these metrics. 

The aggregated accelerometer data was transformed into five-way data, with $n_1=4$ being the seconds, $n_2=7$ being the metrics, $n_3=6$ being the minutes and $n_4=12$ being the hours. The $n^*$ value is 2016, making this data comparable to the $n^*=2401$ and $N=90$ combination in our simulations. The goal of the analysis was to cluster the youth into groups based on their physical activity profiles and determine if these groups agree with their clinical groupings or are capturing additional information about the participants.

\begin{figure}
	\begin{center}
		\includegraphics[width=4in]{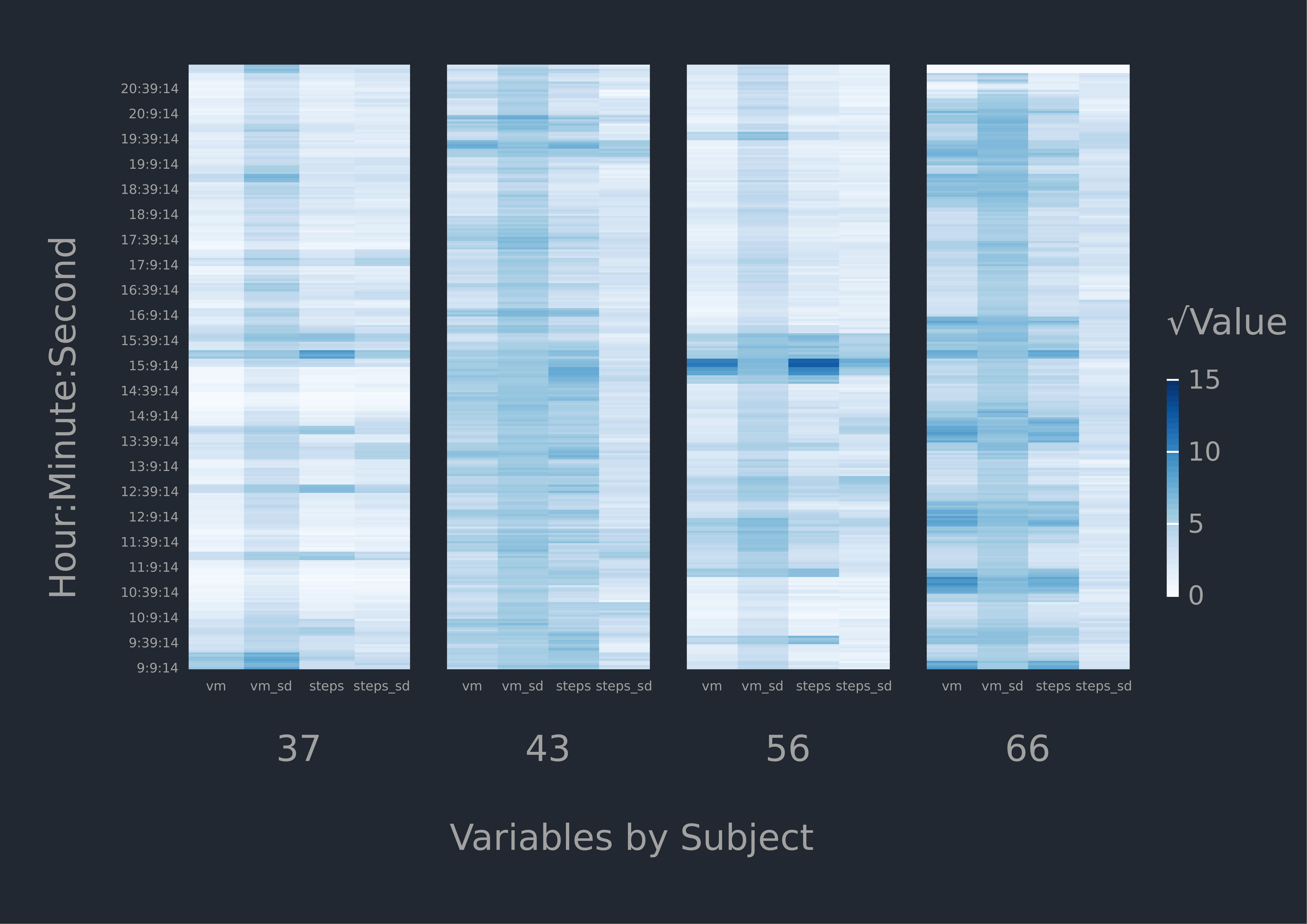}
	\caption{Two columns of a random sample of six $\vecX_{(2),i}$ matrices.}
	\label{fig:c_subj}
\end{center}
\end{figure}

Figure \ref{fig:c_subj} shows the mode 2 matricization of four random participants accelerometer data. It is clear that there is a lot of variation between the youths' activity patterns across the aggregation levels. Participant 37 is moving very little through out the day. Participants~56 and 66 exhibit short bursts of activity, highlighted by the dark blue bands in the VM and steps columns. A pattern most evident in participants 37 and 66, suggests that over the course of a day, VM patterns can be highly variable, as measured by the SD of VM but this variability is not reflected in the intensity of that movement or the number of steps. In general, the youths' movement patterns exhibit short intervals of intense movement and consistent variation in their movement, irrespective of intensity. These intervals occur most often mid-morning and early afternoon.

\begin{figure}
	\begin{center}
		\includegraphics[width=4in]{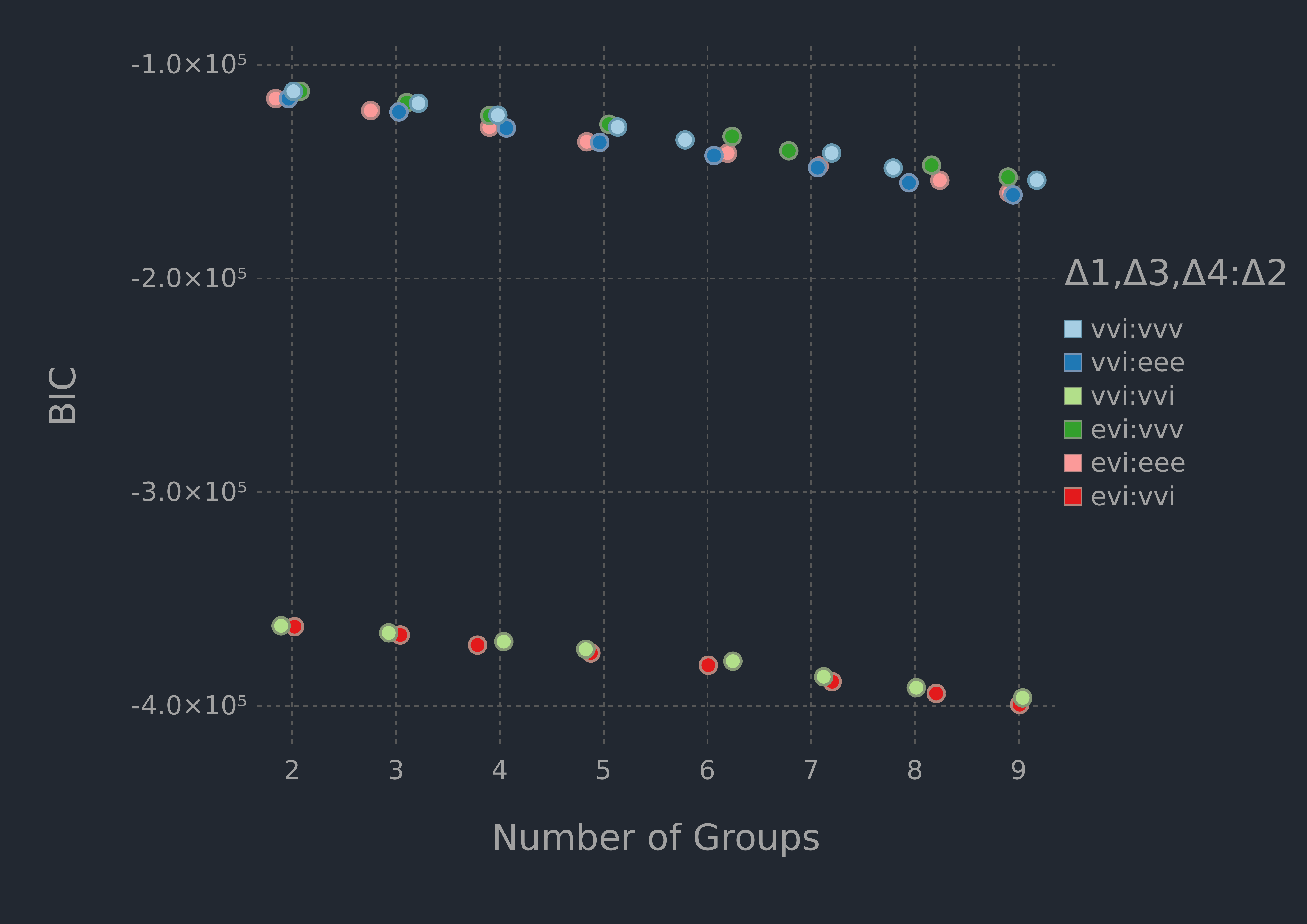}
	\caption{BIC by group size and model type.}
	\label{fig:c_bic}
\end{center}
\end{figure}

When clustering the sample of five-way data, we started by assessing six different mixture models over 9 different group sizes. The models BIC values are summarized in Figure~\ref{fig:c_bic}. The figure's legend is interpreted as follows: the first three letters correspond to the temporal scale model and the last three letters correspond to the scale model used for the accelerometer metrics. In general, a two or three group solution with an unstructured $\vecDelta_{2, g}$ resulted in better clustering solutions, as determined by BIC. The model with the largest BIC was the one with two groups and used the VVI model for the temporal scale dimensions and VVV for the accelerometer metrics. Given the inter-dependencies between the metrics, it is not surprising that the VVI models for $\vecDelta_{2, g}$ fared so poorly. The slight edge shown by the VVV models of $\vecDelta_{2, g}$ over EEE, implies each group has a different pattern of co-variation between the metrics. For the same group size, the temporal scale models had very similar performance for the same $\vecDelta_{2, g}$ model. This suggests the temporal relationships are very similar between groups. None of the models we investigated had a singular scale matrix.

\begin{figure}
	\begin{center}
		\includegraphics[width=4in]{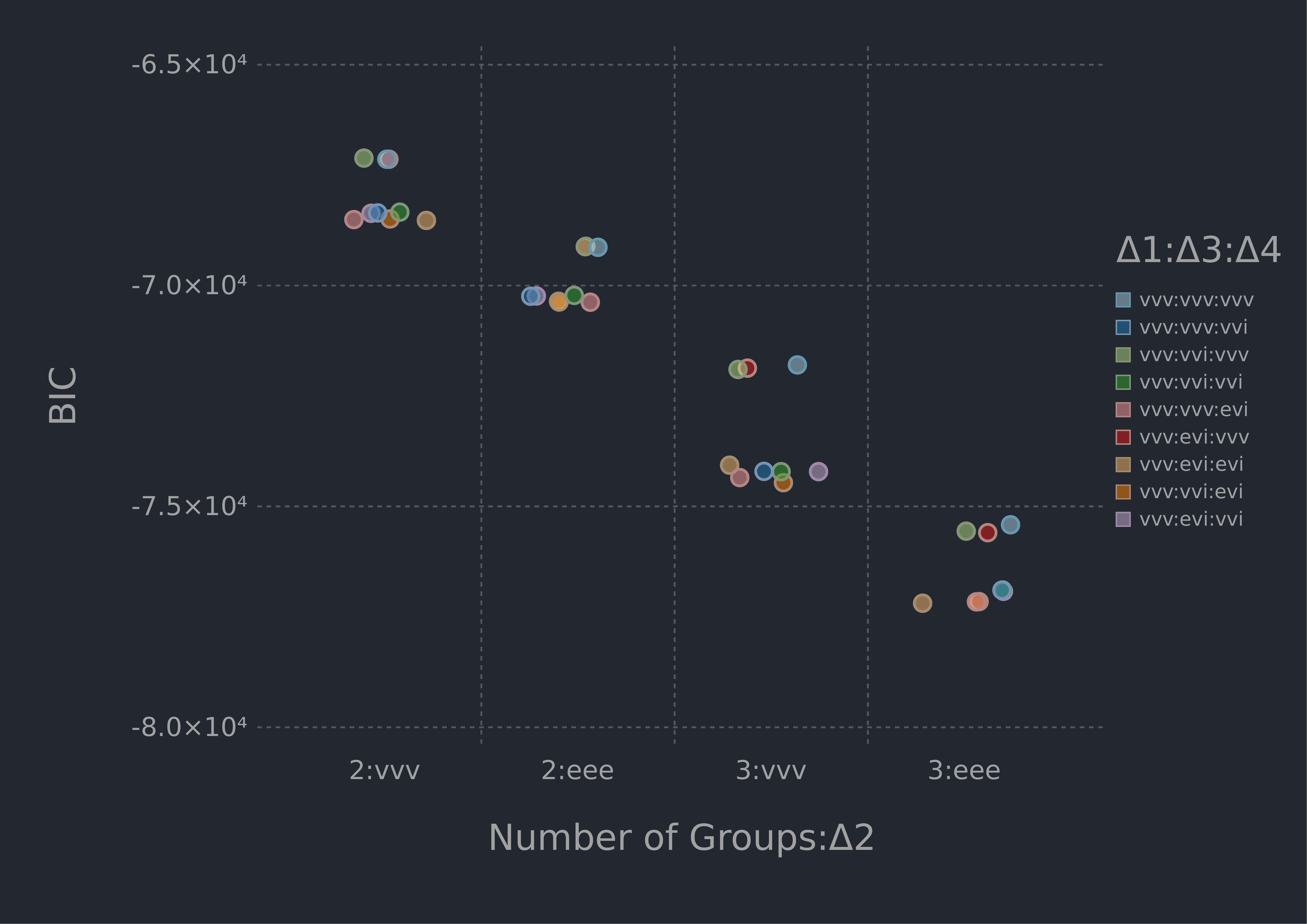}
	\caption{BIC by group size and model type.}
	\label{fig:c_bic2}
	\end{center}
\end{figure}

To further improve on the clustering solution, we focused on two and three group solutions where each scale model could have any of the three appropriate options. The results are displayed in Figure~\ref{fig:c_bic2}, which includes some horizontal jittering to clarify the points position. Models with unstructured scale matrices for the second and hour dimensions are preferred based on BIC. For this sample of five way data, these dimensions do not exhibit patterns of variation consistent with an autoregressive model, despite representing measurements taken over time. Models that use an autoregressive model to model the minutes are consistently preferred. Using an EEE model to model the variation between the accelerometer metrics is not preferred, suggesting the groups do not exhibit the same pattern of variation across the metrics. The VVI temporal model for $\vecDelta_{3, g}$ is marginally better than EVI, implying that the groups have different autoregressive patterns in the minute dimension of their MDAs. It should be noted that even the poorest performing models in Figure~\ref{fig:c_bic2} have much larger BIC values than the best models in Figure~\ref{fig:c_bic}. The final model is a two group solution with the VVI temporal model for $\vecDelta_{3, g}$ and unstructured scale matrices for the other three dimensions. Flexibility in how the variation of each dimension of the MDA is modelled is an important feature of these models and results in better clustering solutions, as determined by BIC. 

\begin{figure}
	\begin{center}
		\includegraphics[width=4in]{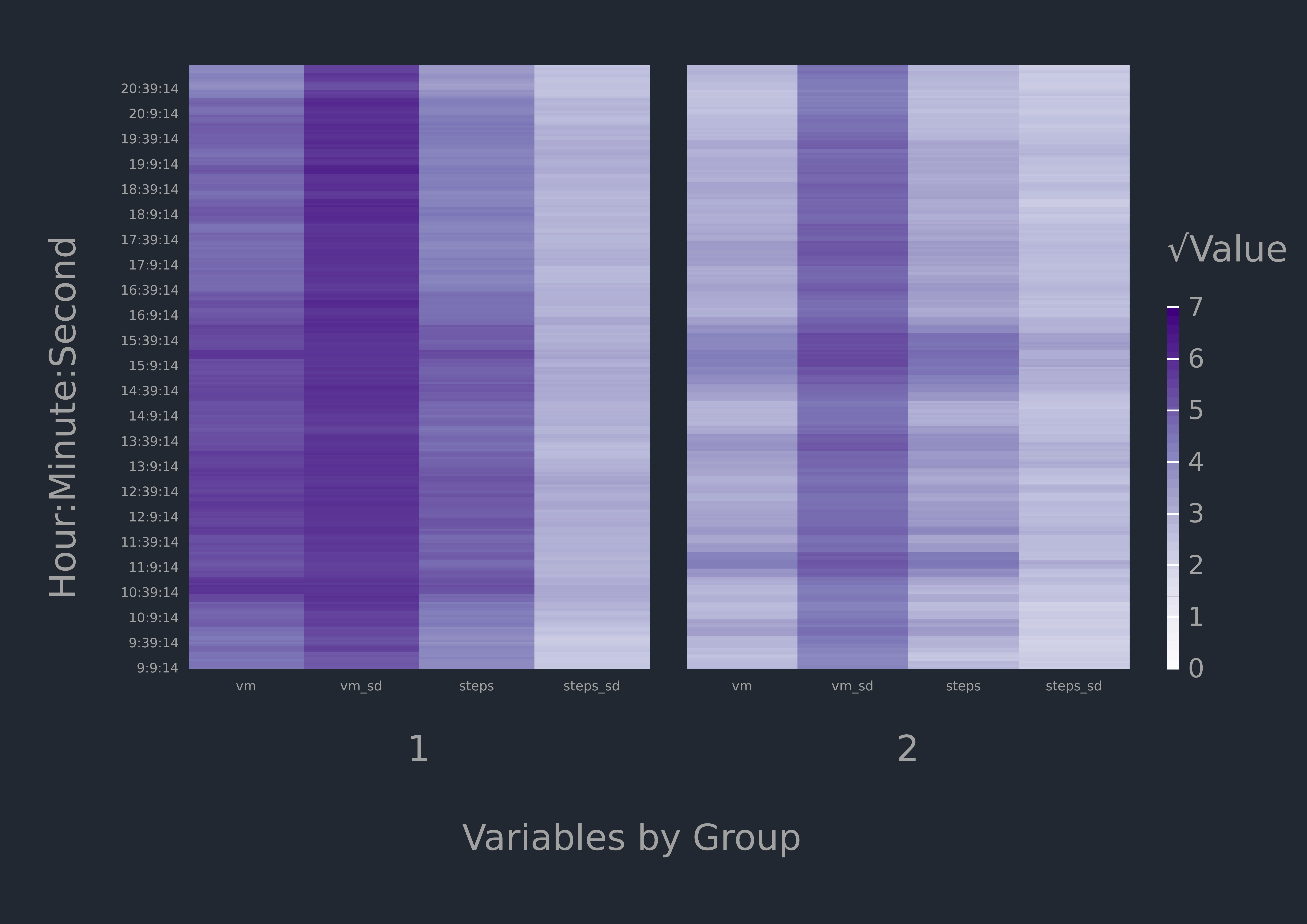}
	\caption{Mean vector magnitude and steps by group. A column of the $\vecM_{(2),g}$ matrices.}
	\label{fig:c_mean}
\end{center}
\end{figure}

Figure \ref{fig:c_mean} plots the mode 2 matricized mean tensor $\vecM_{(2),g}$, which visualizes the mean activity patterns of each group. Group 1 is the group with the most consistently active participants, moving at higher intensities through out the day. Group 2 exhibits brief periods of intense activity in the mid morning and late afternoon, illustrated by the deep purple bands in the VM and steps columns. Both groups have similar mean patterns of variation in their steps. 

\begin{figure}[h]
	\begin{center}
		\includegraphics[width=4in]{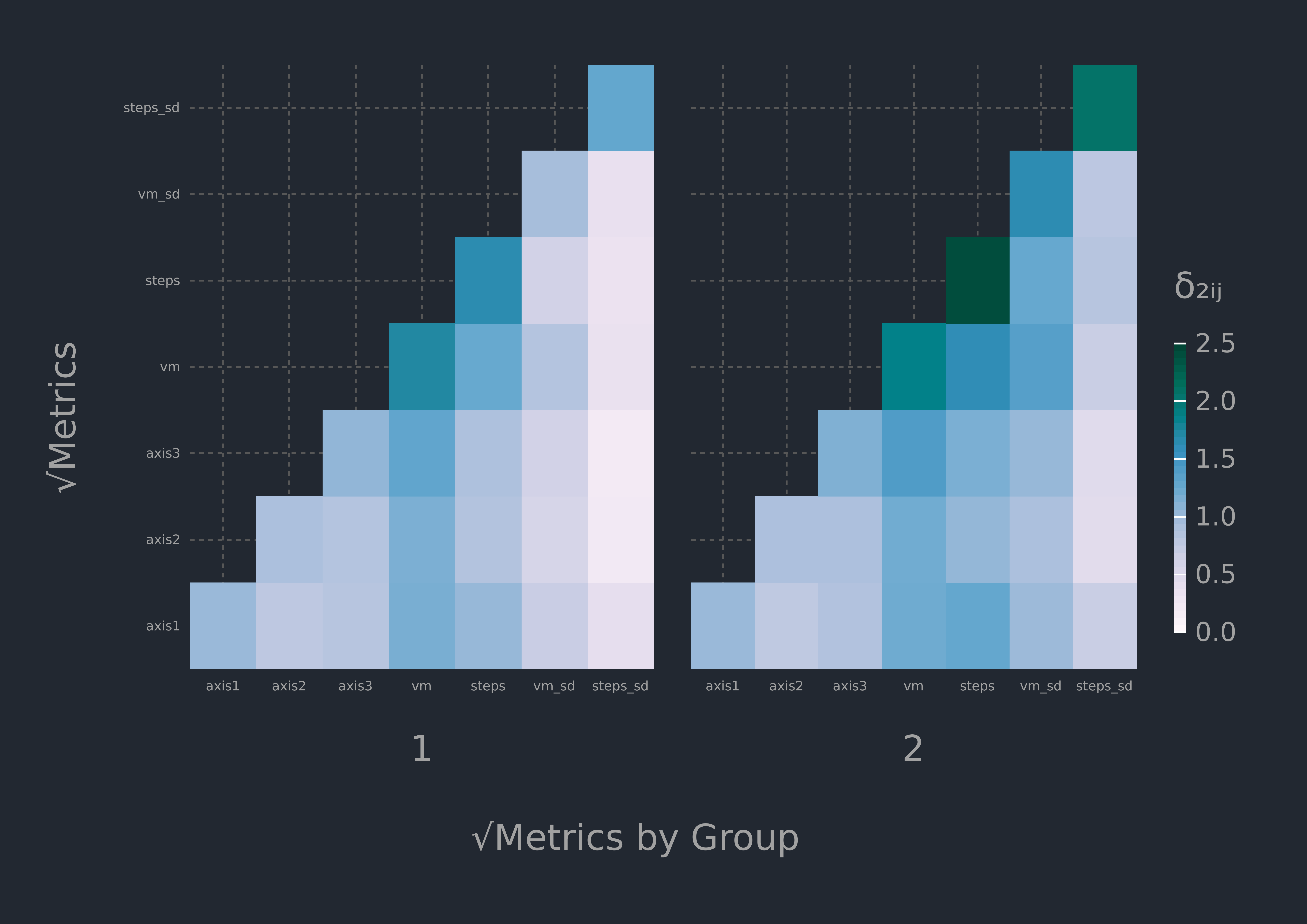}
	\caption{Scale matrix for the accelerometer metrics, $\vecDelta_{2, g}$.}
	\label{fig:c_d2}
\end{center}
\end{figure}

Figure \ref{fig:c_d2} shows the variation of the accelerometer metrics, represented by $\vecDelta_2$, plotted by group. The individual entries of the lower triangular portion of the matrix, $\delta_{2,i,j}$ are plotted as a heatmap. Group 2 has more variation in their metrics, as is evident by the deeper shades of blue and green in its heatmap. This makes sense in light of the participants periodic bouts of intense movement interspersed with longer periods of little activity. Steps have the most variation of the five metrics, followed by VM. As expected, VM covaries with the three axis values and steps covary with axis 1. In group 1, the more active participants, variation in VM (vm\_sd) does not covary with any metric except VM, unlike group 2. Variation in steps (steps\_sd) does not covary with any metric, including steps in group 1.

\begin{figure}[!h]
	\begin{center}
		\includegraphics[width=4in]{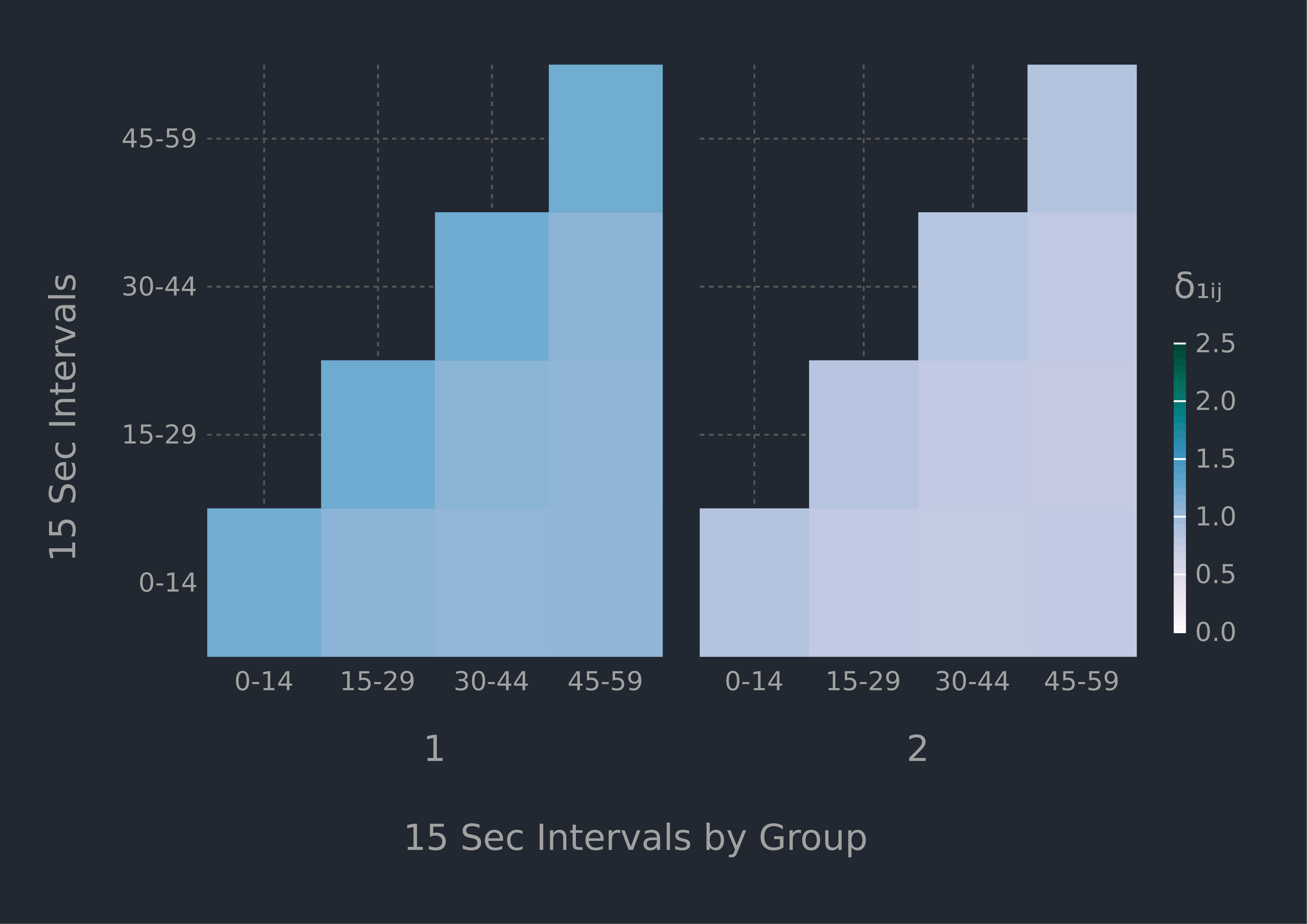}
	\caption{Scale matrix for 15 second intervals, $\vecDelta_{1, g}$.}
	\label{fig:c_d1}
	\end{center}
\end{figure}

Figure \ref{fig:c_d1} plots the variation in the four 15 second intervals, captured by $\vecDelta_1$. Group~1 exhibits more variation at this time scale than group 2. In both groups, the levels of covariation between the intervals remains nearly constant. This is contrary to the VVI and EVI temporal models, that would impose decreasing levels of variation as you move away from the main diagonal. 

\begin{figure}
	\begin{center}
		\includegraphics[width=4in]{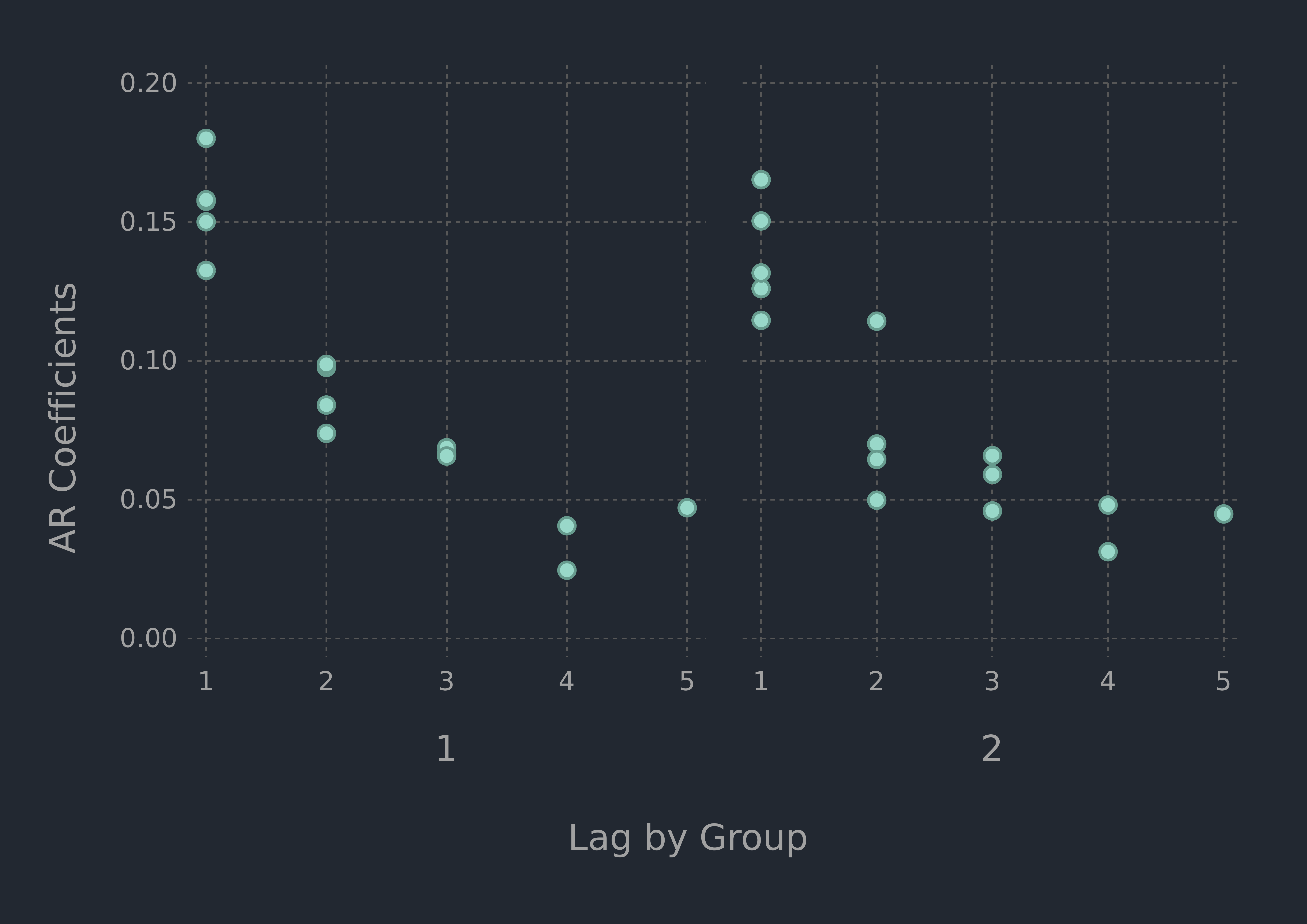}
	\caption{Autogressive coefficients for 10 minute intervals, $\matt_{3, g}$.}
	\label{fig:c_T3}
	\end{center}
\end{figure}

Figure \ref{fig:c_T3} displays the AR coefficients from the $\matt_{3, g}$ matrices derived from the 10 minute intervals. The pattern exhibited by the coefficients are generally decreasing with increasing lag between time points. This is characteristic of true longitudinal data. In general, the magnitude of the early AR relationships are shifted downward in group 2 relative to group~1. This suggests the AR relationships are slightly stronger in the active participants. Because the two sub-plots are nearly interchangeable, it is not surprising how similar the BIC results were for the models that used the VVI and EVI models for this dimension. The two isotropic constraints representing the variation at each time point are nearly identical between the groups ($\delta_{3,1}=0.984, \delta_{3,2}=0.987$) and relatively small in magnitude. This suggests the AR relationships are the important feature differentiating the two cluster groups in this dimension.

\begin{figure}
	\begin{center}
		\includegraphics[width=4in]{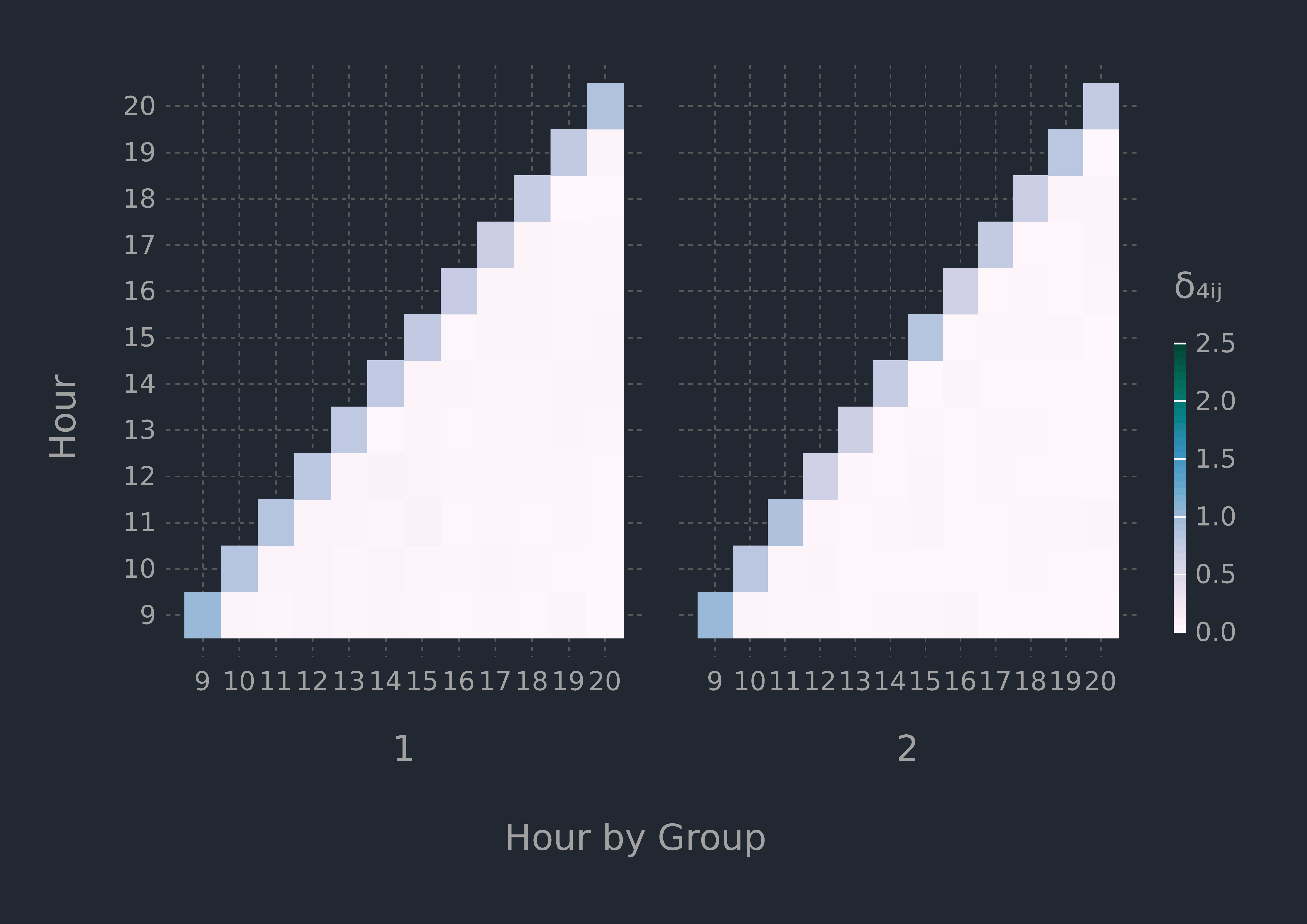}
	\caption{Scale matrix for the hours, $\vecDelta_{4, g}$}
	\label{fig:c_d4}
	\end{center}
\end{figure}

Figure \ref{fig:c_d4} displays the variation in the hours captured by $\vecDelta_{4, g}$. There is little variation in each hour and it remains consistently low through out the day. Covariation is nearly absent between the hours. These patterns are evident in both groups and imply that the hour dimension does not help distinguish between the participants in each cluster group. This aligns with the existing observations that children move in short bursts of activity through out the day and these patterns are more evident when analyzing higher resolution data at multiple time scales \citep{baquet2007, rowlands2008}. 


A cross-tabulation of the clusters (based on MAP estimates) versus the medical diagnosis groupings is given in Table~\ref{tab:ct_c}. Each cluster represents roughly half the participants. The control and T1D participants are evenly distributed between the two groups. Group 1, the active participants, has the majority of the CF youth ($71\%$)  while group 2 has the majority of the IBD ($65\%$) youth. The ARI for the table is 0.0007, indicating that the labels produced by the model are unrelated to the youth's membership in one of the five study groupings. This indicates that within each CIC, there is heterogeneity in the youths physical activity profiles. This heterogeneity can be used by clinicians to inform the activity recommendations for low fit youth based on their high fit counterparts that share the same CIC.
\begin{table}
	\centering
	\caption{Cross-tabulation of the clusters (based on MAP estimates) versus the medical diagnosis groupings.}\label{tab:ct_c}
	\begin{tabular}{lrrrrrr}
		\hline
		\textbf{Cluster} & \textbf{Control} & \textbf{CF} & \textbf{IBD} & \textbf{JIA} & \textbf{T1D} & \textbf{Total(\%)}  \\ \hline
		1 &7 &10 &6  &12 &8 &43(52) \\ 
		2 &7 &4  &11 &9  &9 &40(48) \\ \hline
	\end{tabular}
\end{table} 

\begin{figure}[h]
	\begin{center}
		\includegraphics[width=3.55in]{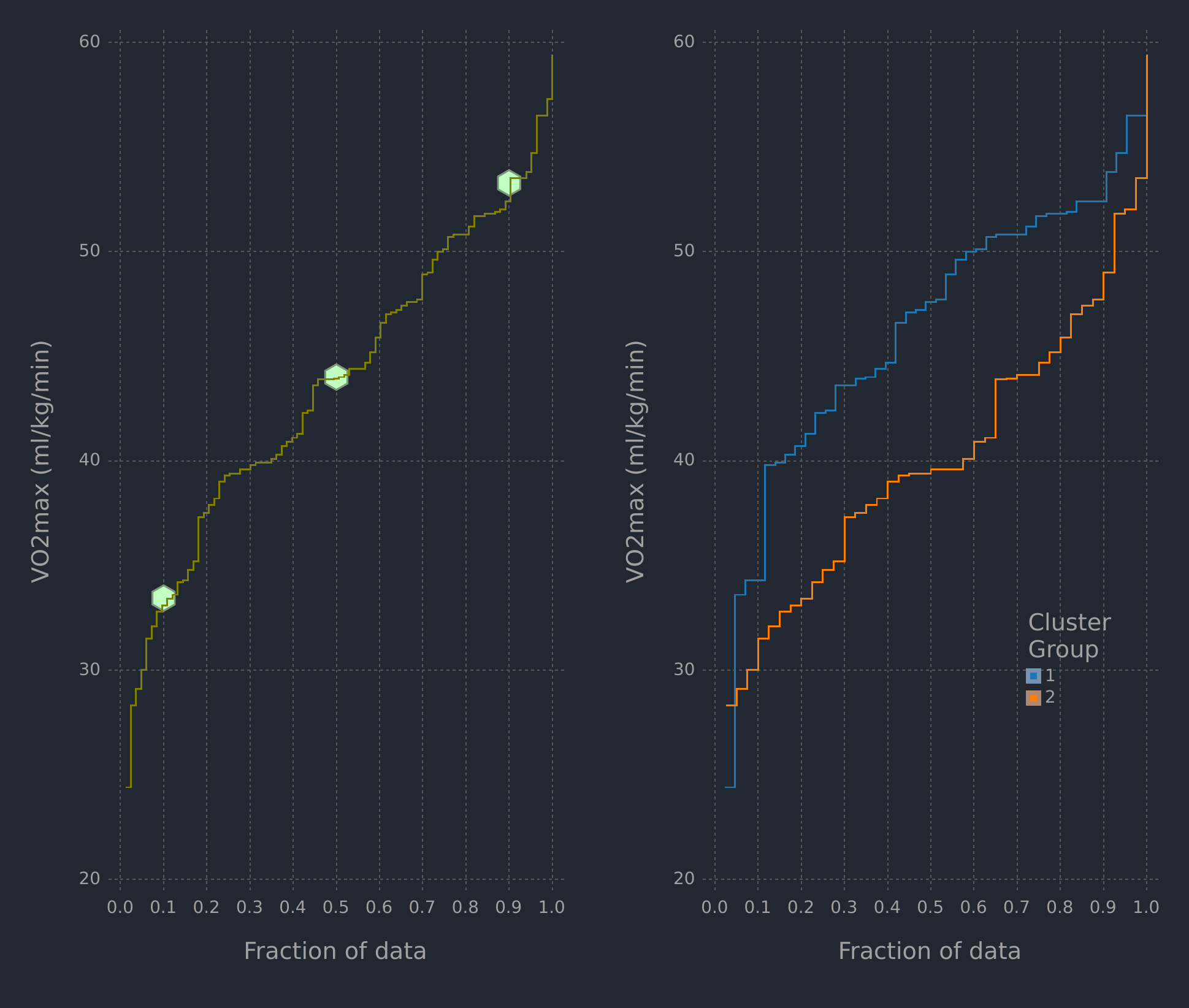}
	\caption{Empirical cumulative distribution plots of VO2max overall and by cluster group.}
	\label{fig:vo2}
	\end{center}
\end{figure}

The results of the finite mixture of MDAs model could be useful in a number of ways. Beyond characterizing the covariation in each mode of the samples MDAs, the group labels can be used in conjunction with traditional statistical models. We use them as one of four predictors in a quantile regression characterizing the relationship between predictors and aerobic fitness, defined as maximal oxygen uptake or VO2max. VO2max is the gold-standard measurement of cardiorespiratory fitness and is the maximum rate of oxygen consumption measured during incremental exercise \citep{vo2max}. Our VO2max values are expressed as a relative rate, millilitres of oxygen per kilogram of body mass per minute (ml/kg/min) to account for differences in body size between participants in the sample. The distribution of VO2max is illustrated in figure \ref{fig:vo2}. The left hand panel illustrates the empirical cumulative distribution plot for all $83$ participants. The green hexagons indicate the following quantiles ($\tau$), used in our regression model: $10\text{th} = 33.4$, median/$50\text{th} = 44.0$ and the $90\text{th} = 53.3$. The $10\text{th}$ and $90\text{th}$ quantiles represent the youth in our sample with low and high cardiorespiratory fitness. The right hand panel illustrates the empirical cumulative distribution lines for each cluster group. Group 1, the active youth, have larger VO2max values across the quantiles which aligns with our expectation that the more active youth would have better cardiorespiratory fitness. This indicates our finite mixture model effectively captured the salient information about the participants physical activity from the accelerometer data.

At the time of writing, we have access to the participants age, sex, study arm and VO2max values. Basic summary statistics by cluster group are listed in Table~\ref{tab:bl}. Group~1 is slightly younger, predominantly male and has higher cardiorespiratory fitness than group~2.  Cluster~2 is older and predominantly female, suggesting daily patterns of physical activity are different for each sex in our sample. This is consistent with existing literature demonstrating that girls engage in less physical activity compared with boys, starting as early as the preschool years, in early childhood, as well as in adolescence \citep{proudfoot2019, farooq2020}. The age of the CF (median: 11.2)  and the IBD (median: 14.3) participants can help explain their predominance in group 1 and group 2, respectively. Age, and specifically the post-pubertal period, is inversely related to physical activity \citep{nader2008, farooq2020}.
\begin{table}
	\centering
	\caption{CHAMPION Summary statistics by cluster.}
	\label{tab:bl}
	\begin{tabular}{cccccccccc}
		\toprule
		\textbf{Cluster} && \multicolumn{2}{c}{\textbf{Age}(years)} && \multicolumn{2}{c}{\textbf{Females}} && \multicolumn{2}{c}{\textbf{VO2max}(ml/kg/min)} \\
		\midrule
		{} && Mean & SD && $n$ & \% && Mean & SD \\ 
		1 && 11.8 & 2.63 && 14 & 33 && 47.0 & 6.82  \\ 
		2 && 14.2 & 2.05 && 29 & 73 && 40.6 & 7.10  \\ 
		\bottomrule
	\end{tabular}
\end{table} 

\begin{figure}
	\begin{center}
		\includegraphics[width=4in]{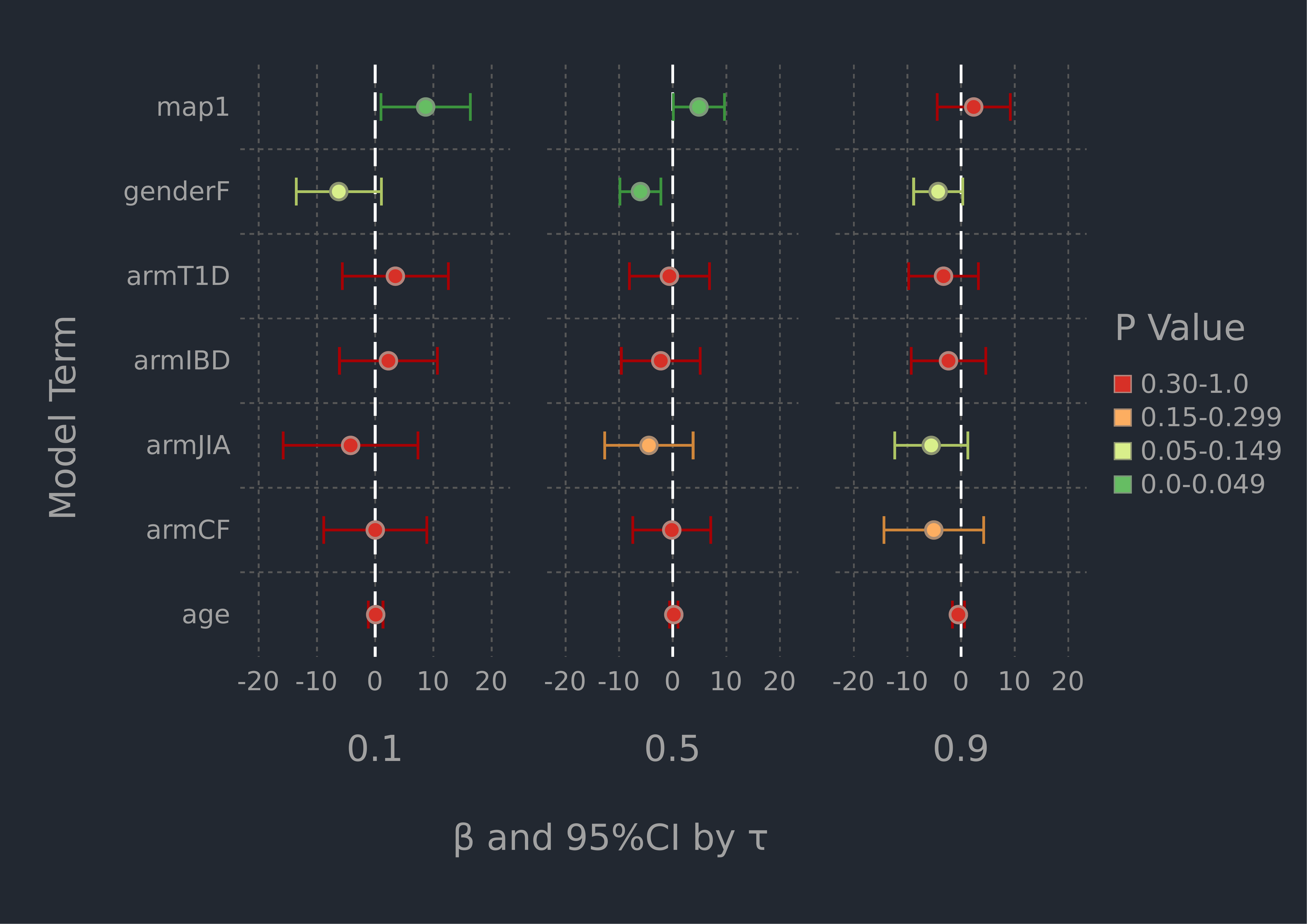}
	\caption{Quantile regression coefficients.}
	\label{fig:qr_coef}
	\end{center}
\end{figure}

The Julia package \texttt{RCall} is used to interoperate with \textsf{R} version 3.6.3 \citep{Rcitation} and the \texttt{quantreg} package, version 5.61 \citep{quantreg} to conduct the quantile regression. The model coefficients $\beta$, their $95\%$ confidence intervals, and p-values are summarized in Figure~\ref{fig:qr_coef}. We did not include interactions between the models main effects due to the limited sample size. The $\beta$ values for the study arms are relative to the control group. For the less fit participants, being in the active group vs the inactive group has a relatively large and statistically significant effect on VO2max at the $5\%$ level --- $\beta=8.68$, $95\%$ CI $(0.88, 16.36)$ --- while being female vs male, trends towards a decreased VO2max --- $\beta=-6.25$, $95\%$ CI $(-13.56, 1.06)$. When controlled for the other predictors, the effect of the study arm and age do not have an important relationship with VO2max. The effect of age on VO2max is likely muted by the inclusion of our cluster groups, which differ in age. Similar trends are evident at the median VO2max, with both cluster group --- $\beta=4.89$, $95\%$ CI $(0.11, 9.67)$  and sex --- $\beta=-6.03$, $95\%$ CI $(9.84, -2.21)$ being statistically significant. At the $90\text{th}$ quantile, the high fit youth do not have any statistically significant relationships between the predictors and VO2max. The same trends in MAP and sex are evident here and being in the control group may confer some increase in cardiorespiratory fitness.  

In summary, our finite mixture of MDAs model found two groups in the five-way accelerometer data from the CHAMPION study. One group represents participants that are consistently active through out the day and another that exhibits long periods of light activity interspersed with intervals of intense activity. The groups are differentiated by their mean activity profiles, variation at the level of 15 second intervals and between their accelerometer metrics and the strength of the AR relationships between their 10 min intervals. These groupings do not agree with the participants clinical diagnoses, suggesting the accelerometers are capturing information not included in these diagnoses. The group labels produced by our model were good predictors of the participants cardiorespiratory fitness, as measured by VO2max, when the participants had low to median fitness.

\section{Discussion}

Physical activity data, as measured by accelerometers, was conceptualized as a sample of five-way data, where each participant has an order-$4$ MDA made up of their accelerometer data aggregated over different nested time scales. A finite mixture of MDAs approach is introduced for clustering a sample of MDA data. These models are innovative because they allow MDA data to be analyzed in its natural form, without the need to transform it to meet the limitations of existing model-based clustering methods. Our model provides scientifically relevant parameters for each group, characterizing the mean MDAs, the variation in each dimension of the MDA and labels that can be used in conjunction with more traditional statistical methods. These parameters can help non-statisticians make practical decisions and guide their scientific messaging around the MDA sample being analyzed.

Clinical data, like the data from the CHAMPION study, often includes non-continuous covariates. These covariates can be informative (e.g., sex, disease status) and should be included in the clustering solutions when ever possible. One way to do this in models like the one describe herein, would be to model the mean of each mixture component with a linear model \citep{mcnicholas2012} that incorporates these covariates. The matrix $\vecM_{(1),g}$ could be modelled using a matrix-variate regression model that models matrix valued responses \citep{ding2018}. Alternatively, the MDA $\tenM_g$ could be modelled using a tensor-variate regression model \citep{li2017}. These models have a one-step estimator which is computationally quick and has competitive finite sample performance compared to iterative solutions, making them an attractive target to integrated into the EM algorithm described in Section~\ref{sec:vvv}.  

The quality of the metrics included in the MDAs is important for any analysis. When dealing with data collected by sensors, like accelerometer data, it is common to create aggregated summaries in pre-defined time intervals. This can potentially cause a substantial loss of information about the signal in the data and alter relationships between accelerometer-measured physical activity and health outcomes \citep{bailey2013, rowlands2019}. One way to overcome this is to include metrics that capture important features of the data. In the context of accelerometer data, some effort has been made to classify human physical activity using raw accelerometer data \citep{lyden2014, rowlands2018, rowlands2019, kos2020}. These features or labels tend not to be continuous data and would not be directly applicable to the finite mixture of MDAs model described here. Alternative continuous metrics derived from raw accelerometer data are described in \cite{karas2019} and could likely improve the clustering solutions. The vector magnitude count or mean amplitude deviation \citep[VMC/MAD;][]{vaha2015} has been shown to be accurate at classifying the magnitude of bipedal motion. The activity index \cite[AI;][]{bai2016}, is a metric that combines the three within-axis standard deviations of the raw accelerometer data. AI is additive and rotationally invariant making it easy to aggregate and robust to
rotations or tilts in the accelerometer placement during the participants free-living activities. The AI has been shown to be able to distinguish between different types of physical activity across a range of intensity levels and is a good predictor energy expenditure during the accelerometers wear time.

This work could be extended in several methodological directions. The regularization mentioned in Section~\ref{sec:imp} is often done implicitly in software implementations such as {\tt scikit-learn}'s GaussianMixture function, written in Python. The value of the regularization parameter, $\epsilon$ could be tuned. The larger it is, the further the model results are from the true solution. In supervised learning, $\epsilon$, could be chosen using cross-validation. In the unsupervised context, choosing the optimal value of $\epsilon$ could be done via model selection using the BIC, similar to \cite{bouveyron2014} approach to selecting $\lambda$, the LASSO regularization parameter in their latent mixture model that selects discriminative variables in high dimensional clustering problems.

We expect clustering MDA data would benefit from some form of dimension reduction. The VVI model that constrained the eigen-decomposition of $\matdel_{g, 2}$, had $n_d$ vs. $n_d^2$ parameters in the EEE and VVV models. Despite this large reduction in free parameters, the VVI model was not able to improve the performance of the mixture model, as measured by BIC. Having scale matrices that are unstructured was an advantage in our analysis in section \ref{sec:champ}. We expect this to be true when doing dimension reduction as well. In this vein, a finite mixture of MDA factor analyzers could be developed, and can be viewed as an extension of the work of \cite{tang2013} and \cite{gallaugher2018msmvbf}.


%
%
%
%
%

\bigskip
\appendix
\section{Mathematical Details}

\subsection{Equation \texorpdfstring{\ref{eq:trace2}}{Trace}}\label{app:A}

\begin{align*}
\nonumber
\tr&\left[\vecDelta_1^{-1}\vecXbrev_{(1)}\T\bigotimes_{d=2}^D\vecDelta_{d}^{-1}\vecXbrev_{(1)}\right] = \tr\left[\vecDelta_1^{-1}\vecXbrev_{(1)}\T\vecDelta_{2}^{-1}\otimes \bigotimes_{d=3}^D\vecDelta_{d}^{-1}\vecXbrev_{(1)}\right]\\
\nonumber
&=\tr\left[\vecDelta_1^{-1}\vecXbrev_{(1)}\T
(\ident_{n_2}\vecDelta_{2}^{-1}\ident_{n_2})\otimes\left(\bigotimes_{d=3}^D\vecDelta_{d}^{\mhalf}\ident_{n^{*}_{3:D}}\bigotimes_{d=3}^D\vecDelta_{d}^{\mhalf}\right)\vecXbrev_{(1)}\right]\\
\nonumber
&=\tr\left[\vecDelta_1^{-1}\vecXbrev_{(1)}\T
\left(\ident_{n_2}\otimes\bigotimes_{d=3}^D\vecDelta_{d}^{\mhalf}\right)
\left(\vecDelta_{2}^{-1}\sum_{j=1}^{n^{*}_{3:D}}\evec{}{j}\evec{\top}{j}\right)
\left(\ident_{n_2}\otimes\bigotimes_{d=3}^D\vecDelta_{d}^{\mhalf}\right)\vecXbrev_{(1)}\right]\\
\nonumber
&=\sum_{j=1}^{n^{*}_{3:D}}\tr\left[\vecDelta_1^{-1}\vecXbrev_{(1)}\T
\left(\ident_{n_2}\otimes\bigotimes_{d=3}^D\vecDelta_{d}^{\mhalf}\evec{}{j}\right)
\vecDelta_{2}^{-1} \left(\ident_{n_2}\otimes\evec{\top}{j}\bigotimes_{d=3}^D\vecDelta_{d}^{\mhalf}\right)\vecXbrev_{(1)}\right]\\
&=\sum_{j=1}^{n^{*}_{3:D}} \tr\left[\vecDelta_1\inv \vecX_{(1), j}\T \vecDelta_{2}\inv \vecX_{(1), j}\right],
\end{align*}

\subsection{Equations \texorpdfstring{\ref{eq:vvi_s1}}{VVI Score 1} and \texorpdfstring{\ref{eq:vvi_s2}}{VVI Score 2}}\label{app:B}

Starting with the determinant term in \texorpdfstring{\eqref{eq:ll1}}{Log Likelihood 1} and the trace in \texorpdfstring{\eqref{eq:ll2}}{Log Likelihood 2}, we rearrage the terms related to $\matdel_{g,1}$:
\begin{align*}
	\nonumber \text{I} && \text{II} \\
	-\frac{n^*}{2}\sum^G_{g=1}n_g\sum_{d=1}^D\frac{1}{n_d}\log(|\matdel_{g,d}|)&& -\frac{1}{2}\sum^G_{g=1}\sum_{i=1}^N z_{ig}\sum_{j=1}^{n^{*}_{3:D}}\tr[\matdel_{g,1}\inv\vecX_{(1),g,i,j}\T\matdel_{g,2}\inv\vecX_{(1),g,i,j}]\\
	-\frac{n^*}{2n_1}\sum^G_{g=1}n_g\log(|\matdel_{g,1}|) && -\frac{1}{2}\sum^G_{g=1}\tr \bigg[\sum_{i=1}^N z_{ig}\sum_{j=1}^{n^{*}_{3:D}}\matdel_{g,1}\inv\vecX_{(1),g,i,j}\T\matdel_{g,2}\inv\vecX_{(1),g,i,j}\bigg]\\
	\dots && -\frac{1}{2}\sum^G_{g=1}\tr \bigg[\matdel_{g,1}\inv\sum_{i=1}^N z_{ig}\sum_{j=1}^{n^{*}_{3:D}}\vecX_{(1),g,i,j}\T\matdel_{g,2}\inv\vecX_{(1),g,i,j}\bigg]
\end{align*}

We define:
\begin{itemize}
	\item $\matlam_{g,1} = \frac{1}{n_g}\sum_{i=1}^N z_{ig}\sum_{j=1}^{n^{*}_{3:D}}\vecX_{(1),g,i,j}\T\matdel_{g,2}\inv\vecX_{(1),g,i,j}$
	\item $\vecA_{g,1} = \matt_{g,1}\T\delta_{g,1}\inv\ident_{l_1 \times l_1}\matt_{g,1}$
\end{itemize} 

\begin{align*}
	\text{I} && \text{II} \\
		-\frac{n^*}{2n_1}\sum^G_{g=1}n_g\log(|\vecA_{g,1}\inv|) && -\frac{1}{2}\sum^G_{g=1}n_g\tr \big[\matdel_{g,1}\inv\matlam_{g,1}\big] \\
	\frac{n^*}{2n_1}\sum^G_{g=1}n_g\log(|\delta_{g,1}\inv\ident_{n_1 \times n_1}|) && -\frac{1}{2}\sum^G_{g=1}n_g\tr \big[\matt_{g,1}\T\delta_{g,1}\inv\ident_{n_1 \times n_1}\matt_{g,1}\matlam_{g,1}\big] \\
	\frac{n^*}{2n_1}\sum^G_{g=1}n_g\log[(\delta_{g,1}\inv)^{n_1}] && -\frac{1}{2}\sum^G_{g=1}n_g\delta_{g,1}\inv\tr\big[\matt_{g,1}\matlam_{g,1}\matt_{g,1}\T\big] \\
	\frac{n^*}{2}\sum^G_{g=1}n_g\log(\delta_{g,1}\inv) && \dots
\end{align*}

The $\mathcal{Q}$ function is now
\begin{equation*}
\mathcal{Q}(\varthet) = C + \frac{n^*}{2}\sum^G_{g=1}n_g\log(\delta_{g,1}\inv) -\frac{1}{2}\sum^G_{g=1}n_g\delta_{g,1}\inv\tr\big[\matt_{g,1}\matlam_{g,1}\matt_{g,1}\T\big]
\end{equation*}

The first score function, we take the partial derivative of $\mathcal{Q}(\varthet)$ w.r.t to $\delta_{g,1}\inv$: 

\begin{equation*}
\pd{\delta_g\inv}\mathcal{Q}(\varthet) \Rightarrow 0 + \frac{n^{*}}{2}\sum^G_{g=1}n_g[\delta_{g,1}\inv]\inv - \frac{1}{2}\sum^G_{g=1}n_g\tr[\matt_{g,1}\matlam_{g,1}\matt_{g,1}\T]
\end{equation*}

\begin{align*}
	\frac{n^{*}}{2}\sum^G_{g=1}n_g\delta_{g,1} - \frac{1}{2}\sum^G_{g=1}n_g\tr[\matt_{g,1}\matlam_{g,1}\matt_{g,1}\T] &= 0 \\
    \sum^G_{g=1}\frac{n_g}{2}\left(n^{*}\delta_{g,1} - \tr[\matt_{g,1}\matlam_{g,1}\matt_{g,1}\T]\right) &= 0 
\end{align*}

The first score function is $\text{S}_1(\delta_{g,1}, \matt_{g,1}) = \frac{n_g}{2}(n^{*}\delta_{g,1} - \tr[\matt_{g,1}\matlam_{g,1}\matt_{g,1}\T])$ and we can simplify and solve for $\delta_{g,1}$ to get $\delta_{g,1} = \frac{1}{n^{*}}\tr[\matt_{g,1}\matlam_{g,1}\matt_{g,1}\T]$.

For the second score function, we take the partial derivative of $\mathcal{Q}(\varthet)$ w.r.t to $\matt_{g,1}$ using the matrix differential approach of \cite{magnus2019} :

\begin{align*}
\pd{\matt_g}\mathcal{Q} \Rightarrow &-\frac{1}{2}\sum^G_{g=1}\frac{n_g}{\delta_g}\left(\tr\left[\text{d}\matt_{g,1}\matlam_{g,1}\matt_{g,1}\T \right] + \tr\left[\matt_{g,1}\matlam_{g,1}(\text{d}\matt_{g,1})\T \right] \right)\\
&-\frac{1}{2}\sum^G_{g=1}\frac{n_g}{\delta_g}\left(\tr\left[\matlam_{g,1}\matt_{g,1}\T\text{d}\matt_{g,1} \right] + \tr\left[\matlam_{g,1} \matt_{g,1}\T\text{d}\matt_{g,1} \right] \right)\\
&-\frac{1}{2}\sum^G_{g=1}\frac{n_g}{\delta_g}\left(2\tr\left[\matlam_{g,1}\matt_{g,1}\T\text{d}\matt_{g,1} \right] \right) \Rightarrow -\sum^G_{g=1}\frac{n_g}{\delta_g}\left(\vecc\left(\matlam_{g,1}\matt_{g,1}\T\right)\T\vecc\left(\text{d}\matt_{g,1}\right) \right) \\
\end{align*}

The second score function is $\text{S}_2(\delta_{g,1}, \matt_{g,1}) =-\frac{n_g}{\delta_{g,1}} \matt_{g,1}\matlam_{g,1} $

\bibliographystyle{chicago}
\bibliography{tait_arxiv} 


\end{document}